%% file: main.tex
\newif\iflncs
\newif\ifnotes

\newif\ifblind\blindfalse

\notesfalse
\lncsfalse

\iflncs
\RequirePackage{amsmath}
\documentclass{llncs} 
\usepackage{StyleSheets/template_lncs}
\usepackage{StyleSheets/global}
\else
\documentclass[runningheads, 11pt]{article}
\usepackage{times}
\usepackage{physics}
\usepackage{StyleSheets/template}
\usepackage{StyleSheets/global}
\fi

\begin{document}

\title{Using the Planted Clique Conjecture for Cryptography:\\ \ \\
Public-Key Encryption from \\ Planted Clique and Noisy $k$-$\mathsf{LIN}$ Over Expanders}
 
\date{}
\author{Riddhi Ghosal \thanks{\texttt{riddhi@cs.ucla.edu}} \\ UCLA \and Isaac M. Hair \thanks{\texttt{hair@ucsb.edu}} \\ UCLA, UCSB \and Aayush Jain \thanks{\texttt{aayushja@andrew.cmu.edu}} \\CMU \and Amit Sahai \thanks{\texttt{sahai@cs.ucla.edu}} \\UCLA}
\iflncs
\institute{}
\else
\fi

\maketitle
\noteswarning
\setcounter{page}{1}
\input{abstract}
\newpage



\input{intro_2.0}

\input{Prelims}
\input{assumptions}
\input{PKE}

\input{Properties}

\input{Evidence}
\input{Search}
\pagebreak
\bibliographystyle{alpha}
\bibliography{Bibliography/abbrev3,Bibliography/crypto,Bibliography/custom}

\end{document}

%% file: abstract.tex
\begin{abstract}
    
   We give a public key encryption scheme that is provably secure against poly-size adversaries, assuming $n^{\log^\alpha n}$ hardness of the standard planted clique conjecture, for any $\alpha \in (0,1)$, and a relatively mild hardness conjecture about noisy $k\mbox{-}\mathsf{LIN}$ over expanders that is not known to imply public-key encryption on its own. Both of our conjectures correspond to natural average-case variants of NP-complete problems and have been studied for multiple decades, with unconditional lower bounds supporting them in a variety of restricted models of computation. Our encryption scheme answers an open question in a seminal work by Applebaum, Barak, and Wigderson [STOC'10].




    \end{abstract}

%% file: intro_2.0.tex
\section{Introduction}\label{sec:intro}

Public-key encryption (PKE)~\cite{DifHel76,rivest1978method} is a fundamental primitive enabling secure communication via open and untrusted channels. Despite the immense success of public-key cryptography as a field, and despite intense interest and effort over nearly five decades, we know of surprisingly few 
sources of hardness
that can be nontrivially utlized in building PKE\footnote{
We say that a hardness conjecture X can be \emph{nontrivially} utilized in building PKE if there exists some other hardness conjecture Y that is not known to imply PKE on its own, and yet we can show that the combination of hardness conjectures X and Y yield PKE.
}.

Furthermore, almost all of these rely on structured hardness  rooted in domains such as number theory, algebra, coding theory or lattices (see, for example~\cite{DifHel76,rivest1978method,STOC:Regev05,C:BFKL93,FOCS:Alekhnovich03}). This is in stark contrast with private-key encryption 
and one-way functions, where any hard-on-average problem satisfying very mild structural requirements suffices (e.g. planted clique problem, random constraint satisfaction problems, and sparse PCA).

The overarching goal of this research is to devise new ways of nontrivially leveraging other sources of hardness to enable public-key encryption (and beyond). One such untapped source of hardness is the domain of statistical inference and average-case complexity. For decades, these areas have excelled at identifying and studying natural average-case problems in domains such as graph theory, combinatorics and machine learning from both algorithmic and hardness perspectives. There are also systematic means to give evidence of hardness in restricted algorithmic models such as sum-of-squares \cite{Shor1985,Nesterov2000,parrilo2000structured,lasserre2001global}, statistical-query model \cite{kearns1998efficient}, low-degree tests framework \cite{barak2019nearly,hopkins2018statistical} and many others. 
Unfortunately, despite an abundant supply of well-studied average-case problems, what we seem to lack are \emph{algorithimic reduction techniques} that can nontrivially leverage these problems to create the kinds of ``trapdoor'' structures that are needed to create public-key encryption.
Can we develop new techniques that would allow us to add to the short list of hardness conjectures that can be nontrivially used to build PKE?

In this work, we humbly add one more source of hardness to this list, namely $n^{\log^{\alpha} n}$ hardness of the classic planted clique conjecture, for any $\alpha \in (0,1)$. We do by developing a new quasi-polynomial-time sparsification technique, that can ``simulate'' a random sparse graph from the point of view of achieving a desired level of expansion. We now elaborate.


\paragraph{Can the planted clique conjecture help us build PKE?}
Let us examine the classic planted-clique conjecture which roughly states that no efficient algorithm can determine, with constant advantage, whether a random graph contains a planted clique of an appropriate size. This conjecture is the oldest \cite{alon98,jerrum92,kuvcera1995expected,karp76} and arguably the most well-studied conjecture in average-case complexity. Moreover, its worst-case counterpart, the max clique problem, is a classic $\mathsf{NP}$-complete problem. Despite decades of research on the planted clique conjecture~\cite{jerrum92,kuvcera1995expected,ITCS:ERSY22,juels2000hiding,applebaum10,TCC:ABIKN23,manurangsi20,hazan2011hard,STOC:AAKMRX07,arora2011computational,berthet2013complexity,brennan2018reducibility,brennan2020reducibility,STOC:HirShi24}, constructing a PKE nontrivially utilizing this conjecture has remained elusive.


A seminal work that kick-started the search of PKE from such average-case complexity problems  was that of Applebaum, Barak and Wigderson \cite{applebaum10} which showed several PKE  schemes from various statistical inference problems. Of particular relevance to this work is a construction recipe that shows how to design a PKE from any local CSP and a \emph{variant} planted graph conjecture that, roughly speaking, posits the hardness of finding a small subgraph planted in a random sparse $k$-regular bipartite graph. 
Indeed, the work of~\cite{applebaum10} 
leaves finding a way to leverage the classic planted clique conjecture directly to build PKE as an open question.

We make significant progress towards this question. In particular, our main result builds a PKE scheme from 
two very well-studied~\cite{alon98,jerrum92,kuvcera1995expected,karp76,ITCS:ERSY22,juels2000hiding,applebaum10,TCC:ABIKN23,manurangsi20,hazan2011hard,STOC:AAKMRX07,arora2011computational,berthet2013complexity,brennan2018reducibility,brennan2020reducibility,STOC:HirShi24,feige2003probable,feige2010finding,deshpande2015finding,dekel2014finding,barak2019nearly,ma2015computational,koiran2014hidden,hajek2015computational,allen2015refute,barak2016noisy,STOC:RagRaoSch17,STOC:FelPerVem15,applebaum2016cryptographic,bogdanov2019xor,STOC:KMOW17} average-case complexity problems, namely:
$n^{\log^\alpha n}$ hardness of the standard planted clique conjecture on graphs with $n$ vertices, for any constant $\alpha \in (0,1)$, 
together with a mild conjecture on the hardness of the noisy $k$-$\mathsf{LIN}$ problem\footnote{
In cryptographic contexts this conjecture is sometimes called Sparse LPN~\cite{FOCS:Alekhnovich03,STOC:IKOS08,bogdanov2019xor,C:DIJL23}, but we will use the terminology 
of noisy $k$-$\mathsf{LIN}$. In some references, only the special case of $k$-$\mathsf{LIN}$ over $\mathbb{F}_2$, i.e. $k$-$\mathsf{XOR}$, is considered.} over expanding graphs. Our $k$-$\mathsf{LIN}$ conjecture when interpreted in the special case of equations over $\mathbb{F}_2$ also gives rise to one of the most studied hard distributions for $\mathsf{MAX}$-$k$-$\mathsf{SAT}$ (e.g. ~\cite{grigoriev01,schoenebeck08,FOCS:FeiKimOfe06,STOC:KMOW17}). 



Indeed, neither of these well-studied conjectures (with parameters that we use) were previously known to imply PKE. To the best of our knowledge, ours is the first work that constructs a PKE making critical use of the standard planted clique conjecture\footnote{That is, the planted clique conjecture on an Erdős–Rényi graph with edge probability $\frac12$}.
At the heart of our work is a new algorithmic technique to leverage this conjecture to create ``trapdoor-friendly'' structures.

We now introduce our hardness conjectures, and go through a brief discussion of their history.

\paragraph{Conjecture 1: The Planted Clique Problem.}
Let $\cG(n,\frac{1}{2})$ denote the standard Erdős–Rényi distribution\footnote{That is: in a graph with $n$ vertices, include an edge between every pair of vertices independently with probability $\frac{1}{2}$.} over random graphs. 
Let $\cG(n,\frac{1}{2},n^\delta)$ denote the distribution that first samples a graph from $\cG(n,\frac{1}{2})$ and then plants a clique on $n^\delta$ randomly selected nodes.

\begin{conjecture}[$n^{\log^\alpha n}$ hardness\protect\footnote{Technically, the conjecture is defined to assert hardness against all $n^{O(\log^\alpha n)}$-size algorithms, but note that  hardness against $n^{\log^\alpha n}$-size algorithms  would guarantee hardness against all
$n^{O(\log^\beta n)}$-size algorithms for any constant $\beta$ satisfying $0 < \beta < \alpha$. So for notational simplicity we slightly abuse notation and  use the name ``$n^{\log^\alpha n}$-hardness of Planted Clique Conjecture'' to mean hardness against all $n^{O(\log^\alpha n)}$-size algorithms, for some $\alpha \in (0,1).$
} of the Planted Clique Conjecture]\label{con:plantedclique}
There exist constants $0<\alpha<1$ and $\delta \in (0, \frac{1}{2})$ such that, for all $n^{O(\log^\alpha n)}$-size algorithms $\cA$,
$$     \Bigg \vert \Pr[\cA(G)=1 \mid G \getsr \cG(n,\frac{1}{2})] -  \Pr[\cA(G)=1 \mid G \getsr \cG(n,\frac{1}{2},n^\delta)] \Bigg \vert \leq \frac{1}{3}.
$$

\end{conjecture}

\begin{remark}
    For conciseness, we refer to the above conjecture simply as ``the planted clique conjecture'' in the body of this paper.
\end{remark}

Note that we only require \emph{the existence} of some constants; for example, $\alpha = 10^{-3}$ and $\delta=10^{-3}$ would suffice.  Even stronger versions have been considered, for example
the ``strongish planted clique conjecture'' of Manurangsi, Rubinstein, and Schramm \cite{manurangsi20}, which posits $n^{o(\log n)}$ hardness for the same problem. 


Planted clique has been rigorously studied in statistical inference and average-case complexity ever since its introduction \cite{alon98,jerrum92,kuvcera1995expected,karp76,ITCS:ERSY22,juels2000hiding,applebaum10,TCC:ABIKN23,manurangsi20,hazan2011hard,STOC:AAKMRX07,arora2011computational,berthet2013complexity,brennan2018reducibility,brennan2020reducibility,STOC:HirShi24,ma2015computational,koiran2014hidden,hajek2015computational}. Many different algorithms have been proposed (e.g. \cite{alon98,feige2003probable,feige2010finding,deshpande2015finding,dekel2014finding}), but none of the currently known algrorithms succeed in determining, with constant advantage, whether there exists a planted clique of size smaller than $o(\sqrt{n})$ in polynomial time. Furthermore, the problem has been very well analyzed in restricted algorithmic models, and is subject to strong lower bounds in these models, including sum-of-squares \cite{barak2019nearly} and other frameworks \cite{jerrum92,chen2023almost,STOC:Rossman08,feldman2017statistical,feige2003probable}, providing rigorous evidence for hardness. The recent work of Hirahara and Shimizu also showed a search-to-decision reduction for planted clique \cite{STOC:HirShi24}, so our conjecture is equivalent to assuming only the hardness of the search version of the problem.

\paragraph{Conjecture 2: Noisy $k$-$\mathsf{LIN}$\ Problem over Expanders.} We now describe the noisy $k$-$\mathsf{LIN}$ problem over expander graphs. This is a particular well-studied instantiation of local CSPs over expander graphs where the predicate is just the linear function. As in standard planted CSPs over expander graphs, the problem is defined with respect to a graph represented by a collection of sets $(S_1,\ldots,S_m)$. Here, $n$ is the number of variables and $m=m(n)$ is the number of clauses (each clause being a sparse linear equation over a subset of the variables), and each set $S_i$ is of size exactly $k=k(n)$ and supported over $\{1, \ldots n\}$. The set system satisfies a property that is called $(\gamma,k,t)$-expansion, which requires that for any collection $\{S_{i\in I}\}$ of size at most $t$, the union of $S_i$ has at least $\gamma\cdot k\cdot |I|$ many distinct variables, where $\gamma$ is a constant bigger than $\frac{1}{2}$. See Definition~\ref{def:expand} for formal details. 

In the context of the noisy $k\mbox{-}\mathsf{LIN}$ problem, one is asked to distinguish between the joint distributions $(\mat{A},\mat b = \mat{A}\mat{s}+\mat{e})$ and $(\mat{A},\mat b = \mat{r})$, where $\mat{A}$ is a sparse matrix in $\bF_p^{m\times n}$ produced by taking an indicator matrix $\mathbf{I} \in \{0, 1\}^{m \times n}$ representing the sets $\{S_1,\ldots,S_m\}$ and replacing each nonzero entry in $\mathbf{I}$ with a random element of $\mathbb{F}_p^*$. Specifically, for all $i$, the $i^{th}$ row of $\mat{A}$ encodes the set $S_i$: we have that $\mat{A}_{i,j} \neq 0$ if $j\in S_i$ and $0$ otherwise. The secret $\mat{s} \in \bF_p^{n}$ and $\mat{r} \in \bF_p^{m} $ are uniformly chosen random vectors of appropriate dimensions, and every entry of $\mat{e}$ is sampled independently from a Bernoulli distribution with  probability $\eta$. Backed by systematic lower bounds over many years \cite{grigoriev01,FOCS:FeiKimOfe06,schoenebeck08,barak2016noisy,STOC:RagRaoSch17,STOC:FelPerVem15,applebaum2016cryptographic,bogdanov2019xor,STOC:KMOW17}, we formalize the following widely believed conjecture:

\begin{conjecture}[Noisy $k$-$\mathsf{LIN}$ Conjecture] \label{con:dwdexpander}
There exists a constant $\gamma \in (\frac{1}{2}, 1)$ such that the following holds.
Let $\alpha, \varsigma$ be any constants with $\alpha \in (0,1)$ and $\varsigma>0$.
Then, for $n$ large enough, prime $p = n^{\Omega(1)}$, $k=\Omega(\log(n))$, we have the following: Let $\mat I \in \{0, 1\}^{m \times n}$ correspond to any $(\gamma, k, 2^{(\log n)^\alpha})$-expanding set system, and let $\mathbf{A} \in \mathbb{F}_p^{m \times n}$ be sampled by replacing each nonzero entry in $\mathbf{I}$ with a random element in $\mathbb{F}_p^*$. It holds that
$$(\mat A, \mat b = \mat{A}\mat s+\mat e),$$ is computationally indistinguishable from $$(\mat A, \mat b = \mat r).$$
We only require indistinguishability with respect to a small constant distinguishing advantage, specifically $\frac{1}{16}$. Above, $\mat s \leftarrow \bF^{n}_p$ and $\mat r \leftarrow \bF^{m}_p$ are uniformly randomly chosen and $\mat e\in \bF^{m}_p$ sampled as:
\[\mathbf{e}_i=\begin{cases} \text{Random element of $\mathbb{F}_{p}$, with probability $\eta= (\log n)^{-\varsigma}$. } \\ 0, \text{ otherwise.} \end{cases}\]
\end{conjecture}

\begin{remark}
    In fact, we can use the search version of the above conjecture, and the matrix $\mat A$ for our PKE scheme will satisfy additional desirable properties\footnote{Most notably, for \emph{every} non-empty set of $\{S_{i\in I}\}$, the union of $S_i$ has at least $|I|^{1-\varepsilon}$ many distinct variables, where $\varepsilon$ here is an arbitrarily small positive constant chosen at time of setup.}. See Section \ref{sec:ass} for details.
\end{remark}

Intuitively, the conjecture becomes more ``secure" for larger values of $k$, and if the expansion threshold $t$ and the error probability $\eta$ are large.  In particular, it is desirable that $t$ is at least $\Omega(\log n)$, and in our case, it is actually much larger, namely $t=2^{\log^{\alpha} n}$. Keeping this  value of $t$  in mind, we can  mentally classify the problem in two regimes when it comes to using noisy $k$-$\mathsf{LIN}$ for designing PKE, based on the noise probability $\eta$.

\begin{itemize}
\item Noise probability $\eta<\frac{O(\log n)}{t}$: In this regime, Applebaum et. al.~\cite{applebaum10} showed how to design a PKE only leveraging the noisy $k\mbox{-}\mathsf{LIN}$ problem for any constant $k\geq 3$. 
Here, the noise probability increases as the number of clauses $m$ gets closer to  $n^{k/2}$, but this causes the expansion parameter $t$ to decrease as well, and in all cases, the noise probability is $\eta<\frac{O(\log n)}{t}$. 
Recall that noisy $k$-$\mathsf{LIN}$ is broken when given approximately $m=n^{k/2}$ clauses~\cite{allen2015refute,barak2016noisy,STOC:FelPerVem15}, in which case no nontrivial expansion is possible. 
Any small non-expanding collection
of sets, in some sense, can serve as a secret key for the scheme. 

Indeed, the $k$-$\mathsf{LIN}$ problem in this regime can be efficiently attacked if given unbounded preprocessing on the graph\footnote{It suffices to use a $t$ sparse vector $\mat{x}$ such that $\mat{x}\mat{A}=0$ to distinguish the samples from random.}. 
The ABW scheme from noisy $k$-$\mathsf{LIN}$ problem alone can be broken given a short advice on the matrix.
This is also related to the so-called ``Alekhnovich" barrier in cryptography for design of encryption schemes from (variants of) the Learning Parity with Noise problem \cite{FOCS:Alekhnovich03}. 



\item In the other regime when $\eta>\frac{\omega(\log n)}{t}$, which is \emph{our} regime, we do not know of any PKE scheme from the noisy $k$-$\mathsf{LIN}$ conjecture alone. This regime differs greatly from the previous one: 
No attacks are known on noisy $k$-$\mathsf{LIN}$ in the second regime, even with unbounded preprocessing on the expander graph.  This is also connected to the notion of efficient non-deterministic refutation for certain CSPs \cite{FOCS:FeiKimOfe06} which only applies to the first regime, and where there has been no progress in the second regime. 

In our setting, we mandate expansion on all sets of size at most $t = 2^{(\log n)^\alpha}$, and the error rate is $\eta = (\log n)^{-O(1)} = (\log t)^{-O(1)} \gg \frac{\omega(\log n)}{t}$. 
At the same time, our construction satisfies the following property with regard to subsets of the clauses: For any set $S$ of $x$ variables, at most $x^{1+\varepsilon} \ll x^{k/2}$ clauses exist that use only variables contained in $S$, where $\varepsilon$ is an arbitrarily small positive constant chosen at time of setup. In particular, this means that the total number of clauses is at most $n^{1+\varepsilon}   \ll n^{k/2}$. 

\end{itemize}

In Section \ref{sec:evidence}, we import some existing results that show unconditional lower bounds supporting Conjecture \ref{con:dwdexpander} in the following restricted models of computation: (1) local tests, (2) linear tests, (3) low degree polynomials, (4) sum of squares algorithms, and (5) $\textnormal{AC}^0$ circuits.

\begin{remark}
An important remark is that there are contrived PAC learning based counterexamples to noisy $\mathsf{k}\mbox{-}\mathsf{LIN}$ in settings where the entire expander graph can be computed by a circuit in a shallow complexity class such as $\mathsf{AC}^0[\oplus]$~\cite{oliveira19}. This is due the fact that the resulting $\mathsf{k}\mbox{-}\mathsf{LIN}$ is approximable by a $\mathsf{AC}^0[\oplus]$ circuit that can be PAC-learned. The work also showed existence of such expander graphs relying on Nisan-Wigderson pseudorandom generator applied to some polynomial sized seed~\cite{nisan1991pseudorandom}.

We note that in our case the expander for $k\mbox{-}$$\mathsf{LIN}$ is sampled from a natural distribution induced by a random $G(n,\frac{1}{2})$, and furthermore multiple steps in our construction will introduce independent randomness for the neighbors of each edge in the expander graph. So the neighbor function simply has too much entropy to be computed by any nontrivial $\mathsf{AC}^0[\oplus]$ circuit.
In fact the current counterexamples \cite{oliveira19} only apply to the setting where the number of samples is superpolynomial with respect to the length of the secret, which is far outside of our setting. Finally, in the setting where $p>2$, and the coefficients for the $k\mbox{-}$$\mathsf{LIN}$ constraints are chosen independently and randomly from $\mathbb{F}_p^*$ (as in our construction), there are no known counterexamples even where the number of samples can be superpolynomial.

\end{remark}

\paragraph{Our Main Result.} We establish the following theorem:

\begin{theorem}[Informal]
If Conjecture \ref{con:plantedclique}  and Conjecture \ref{con:dwdexpander} both hold, then there exists a semantically secure public key encryption scheme.    
\end{theorem}

\subsection{Related Work}
\paragraph{Relationship with~\cite{applebaum10}.} Applebaum, Barak and Wigderson designed several PKE schemes from related assumptions. They have a construction from random noisy $k\mbox{-}\mathsf{LIN}$ for any $k\geq 3$ that is in the regime $\eta<O(\frac{\log n}{t})$. Most relevant to our work is their PKE scheme in the regime when $\eta>\frac{\omega(\log n)}{t}$, which is based on nosiy $k$-$\mathsf{LIN}$\footnote{This PKE is actually quite general, and could work with other local CSPs as well.} for any $k$ and a hidden subgraph assumption over random sparse $k$ regular bipartite graphs. Recently, this assumption was proven to be secure against low degree polynomial algorithms, thanks to the work of Bogdanov, Kothari and Rosen \cite{TCC:BogKotRos23}, however we don't know yet how to reduce the standard planted clique problem to their assumption. Applebaum, Barak, and Wigderson do prove an interesting theorem (see Theorem B.13 in the appendix of \cite{applebaum10}) that connects the planted clique problem to a variant of their bipartite graph assumption via a simple reduction. We now discuss two of the main reasons why Theorem B.13 in~\cite{applebaum10} leads to strictly (much) weaker conclusions than ours.

\begin{itemize}
\item Firstly, the theorem only works with planted clique over significantly sparser graphs, specifically those with edge probability $2^{-\ell}$ where $\ell$ is roughly of the order $\log^{(1-\varepsilon)} n$ for some small constant $\varepsilon > 0$. The standard planted clique problem, on the other hand, sets the edge probability to $\frac12$, and is the most widely studied form of the problem algorithmically.

\item Also, their reduction for very low edge probability works only when considering a weakening of their bipartite subgraph assumption: namely where the associated graph satisfies a weaker property known as unique-neighbor expansion. While it turns out that unique neighbor expansion is enough to establish a few lower bounds for noisy $k$-$\mathsf{LIN}$, it is much weaker than the notion of $(\gamma,k,t)$-expanding set-systems for $\gamma > \frac12$, which is what our result uses.
\end{itemize}

\paragraph{Other Schemes.} There are only a few other works that give PKE from graph-based assumptions. Bogdanov, Kothari and Rosen~\cite{TCC:BogKotRos23} give a variant of one scheme by Applebaum, Barak, and Wigderson~\cite{applebaum10}. In particular, they replace the planted dense subgraph assumption with another non-standard assumption about planted even covers, for which they show hardness against low degree polynomial algorithms. 

Abram, Beimel, Ishai, Kushilevitz, and Narayanan explore using a more aggressive assumption to build PKE with logarithmic-size messages \cite{TCC:ABIKN23}. Hudoba proposed a scheme inspired by the problem of finding a planted clique in a modified random graph \cite{hudoba16}, but this scheme was broken by a polynomial-time key recovery attack \cite{chen24}.

\paragraph{Goldreich PRGs.} Inspired by constraint satisfaction over expander graphs, in 2000 Goldreich proposed low-complexity local candidate one-way functions \cite{Gol00} that were later imagined as candidate low-complexity cryptographic pseudorandom generators ~\cite{CM01,STOC:IKOS08}. The complexity theory and
cryptography communities have jointly developed a rich body of
literature on the cryptanalysis and theory of constant-locality
Boolean PRGs~\cite{Gol00,CM01,FOCS:MosShpTre03,TCC:CEMT09,APPROX:BogQia09,applebaum10,TCC:AppBogRos12,applebaum12,CCC:ODoWit14,applebaum16,STOC:KMOW17,AC:CDMRR18,FOCS:AppKac19}. 
Ever since its introduction Goldreich's PRGs have been conjectured \cite{Gol00,applebaum10,TCC:AppBogRos12,applebaum16} to be secure
on arbitrary expander graphs. 
The cryptanalysis of such PRG candidates bears uncanny resemblance to the algorithmic study of the noisy $k\mbox{-}\mathsf{LIN}$ conjecture (in the regime $p = 2$) over expander graphs due to similarity in structure of the two problems. A typical way to instantiate Goldreich PRG predicates is to consider a function that is a sparse linear function which is XORed with an appropriately chosen non-linear predicate to mimic the noise from the noisy $k\mbox{-}\mathsf{XOR}$ problem \cite{applebaum16}. A 20 year study of Goldreich PRGs in cryptography once again solidifies our confidence in the noisy $k\mbox{-}\mathsf{LIN}$ conjecture.

\subsection{Open Problems}
Of course, the most natural open problem left open by our work is to build PKE assuming \emph{only} some reasonable hardness of the planted clique conjecture. 
However, we believe that the most pressing open problems are: 
\begin{itemize}
\item Can we \emph{further populate the list} of explicit average-case hardness conjectures that can be nontrivially used in building PKE?

\item Can we further populate the list of explicit hardness conjectures that suffice on their own to yield PKE (without needing any ``helper'' hardness conjectures)?

\item
What level of sparse error is needed in the context of ``noisy linear algebra''-type hardness conjectures, like $k\mbox{-}\mathsf{LIN}$, to yield (plausibly sub-exponentially secure) PKE without any additional hardness conjectures?

\item
More broadly, \emph{what are the minimal levels of mathematical structure} that suffice to yield PKE?

\end{itemize}

\section{Technical Overview}
\label{sec:tech}

The technical core of our new PKE construction is the instantiation of an \emph{expanding set system} from the planted clique conjecture. 
Concretely, the set system consists of a set of ``variables" and a set of ``clauses" over those variables.
This set system 
is used to sample ciphertexts that are instances of $k$-$\mathsf{LIN}$ on expanders.

In this overview, we will focus on discussing how we start with a graph $G$ that is either sampled from $\er$ or from $\erp$, and then make a sequence of transformations to it that finally yields this desired set system.

\paragraph{Desiderata.} We begin by stating the properties that we need in order to simultaneously achieve security and decryption correctness: 

\begin{enumerate}
    \item []
    \begin{enumerate}
        \item [\textbf{Goal 1}.] If we start with a graph $G$ sampled from $\cG(n,\frac{1}{2})$, then we want the sets of variables and clauses to satisfy strong expansion properties. In particular, every collection $I$ of up to $t = 2^{\log^\alpha n}$ clauses must include at least $\gamma \cdot k \cdot |I|$ variables, where $\gamma \in (\frac{1}{2}, 1)$ is any fixed constant chosen at the time of construction.
        \item [\textbf{Goal 2}.] If we start with a graph $G$ sampled from $\cG(n,\frac{1}{2},n^\delta)$, then there must exist a set of  $x=\polylog(n)$ clauses that include less than $x$ variables. 
    \end{enumerate}
\end{enumerate}

Specifically, Goal 1 above is needed for security of our PKE scheme; Goal 2 allows us to have efficient decryption for our PKE scheme (see Section~\ref{sec:pke} for details).




\paragraph{An initial naive approach.} The most obvious idea is to use the incidence graph between edges and vertices to define our sets of variables and clauses. More specifically, we could use (i) the set of vertices to represent the variables, along with (ii) the set of edges to represent each clause, where each edge represents a clause depending on the variables for the edge's endpoints. Indeed such a construction would only allow us to achieve an arity of $k=2$ if applied to a standard graph\footnote{This corresponds to noisy $2$-$\mathsf{XOR}$, which is easy to solve.}, which doesn't work for us\footnote{In fact, one of the appendices of Applebaum et. al.~\cite{applebaum10} 
used this naive vertex-edge incidence graph in a reduction.}.

A natural generalization of the above idea is to look at subgraphs of $G$ with larger size. In particular, to obtain a set of clauses with arity $k>2$, we can start by assigning a variable to each $(k-1)$-clique in $G$. Each clause would then correspond to a $k$-clique, with the variables being those for each $(k-1)$-clique inside the $k$-clique. 


If we think of constant $k$, then the number of variables and clauses is polynomial in $n$; however, we do not achieve expansion as per Goal 1. Indeed, in $\er$, we expect there to be cliques of size up to $\Theta(\log n)$, and for each clique of size $x$ in our graph, we will have roughly $\binom{x}{k}$ clauses on only $\binom{x}{k-1}$ variables, which is (very) contracting. In fact, it turns out that if one picks a random set of $x$ vertices, for \emph{any} $x \leq n$, we expect to find $\Theta(\binom{x}{k})$ clauses that contain $\Theta(\binom{x}{k-1})$ variables. The problem here is that the density of small regions in the resulting variable-clause bipartite graph when starting from $\er$ is not substantially different than the density of such regions when starting from $\erp$. We therefore ask the following question:

\begin{center}
    \emph{Can we modify $G$ in a useful way to amplify the difference in local density between the planted clique and the rest of the graph?}
\end{center}

\paragraph{Sparsification.}
Intuitively, our aim is to amplify the density disparity between the portion of the graph inside and outside the planted clique, thereby enabling us to achieve Goals 1 and 2. 
We do so by means of a \emph{vertex}-deletion based sparsification technique. %
We transform $G$ into a larger and sparser graph $H$, and we will now identify variables with $(k-1)$-cliques in $H$, and clauses with $k$-cliques in $H$. We do so as follows:
\begin{enumerate}
    \item Create a set of labels, each of which corresponds to an $\ell$-sized subset of vertices in $G$.
    \item Include each label as a vertex of $H$ with probability $p = n^{-c\ell}$ for some small positive constant $c < \delta$.
    \item Connect two nodes in $H$ with an edge iff their respective labels correspond to two subsets whose union forms a clique (of size at most $2\ell$) in $G$.
\end{enumerate}

This method is called the ``randomized graph product'' and has been used previously to study the hardness of various graph problems \cite{berman92, manurangsi20,STOC:HirShi24}. 
Perhaps the most interesting aspect of this randomized graph product is Step 2, where \emph{almost all} of the labels are discarded. To gain intuition as to why this is helpful to us, let's denote by $\hat{H}$ the graph obtained via the product above, but without Step~2 -- namely, we create a vertex in $\hat{H}$ for every label created in Step~1. Now, observe that corresponding to the clique of size $\Theta(\log n)$ in $G$, we can (efficiently) find a clique in $\hat{H}$ of size  roughly $\binom{\log n}{\ell}$. The fact that this ``huge'' clique in $\hat{H}$ is of size $\omega(\log n)$ turns out to be highly problematic for us: this is because the ``huge'' clique will induce a very high level of compression. Because of the deletion (Step~2), our sparsified graph $H$ does \emph{not} have such a large clique.


\paragraph{Going super-polynomial in $n$ --- the need for large graph $H$.} We pause now to make a crucial observation. Note that if we pick $\ell=O(1)$, then the edge probability in $H$ is still a constant. This implies that even after sparsification, for all $x$, there will exist many subgraphs with $x$ vertices and $\Theta(x^2)$ edges. In other words, small structures in $H$ are still quite dense, and the difference in density as compared to the planted clique is only a constant factor.

To overcome this hurdle, it is imperative that we pick $\ell$ to be superconstant. In particular, we choose $\ell=\log^\alpha n$, for any constant $\alpha \in (0, \frac{1}{2})$. While this does yield an $H$ whose size in slightly superpolynomial in the size of $G$, we now have a very useful property: $H$ has an edge probability of $(\frac{1}{2})^{(\log n)^{\Theta(1)}}$. In fact, it turns out that with these parameters, there exists a cutoff value $t = 2^{(\log n)^{\Theta(1)}}$ such that any subgraph of $H$ with $x \leq t$ vertices will have at most only $x\cdot\polylog(n)$ edges! At the same time, if $G$ had a planted clique, then $H$ still would contain a large planted clique, in which subgraphs of the same size will have $\Theta(x^2)$ edges. 

Since the size of $H$ is superpolynomial in $n$, we need the planted clique conjecture to hold against adversaries running in time \\ $n^{O(\log^{\alpha} n)}$ as stated in Conjecture~\ref{con:plantedclique}. 

\paragraph{Constructing the Incidence Graph.} Our remaining goal is to create a mapping from $H$ to variables and clauses. Drawing on the example from the start of this section, we could have a variable for each vertex and a clause for each edge. Obviously, as observed before, such a construction does not work because the arity is $k=2$, but it also fails for a more interesting reason: large star graphs\footnote{Star graphs of size $k+1$ are complete bipartite graphs whose node sets have size $1$ and $k$.} always exist in $H$, and they correspond to sets of clauses with poor expansion properties. Since every edge in the star graph shares a vertex, this means that each clause corresponding to an edge of the star graph has only one unique variable, and the other variable is shared by all of the other clauses. So the best expansion factor we can hope for is $\gamma \leq \frac{1}{2}$, no matter how aggressive our sparsification is. This violates Goal 1 where we need $\gamma> \frac{1}{2}$.

\paragraph{Looking at $k$-cliques.}  To get some intuition, let us extend the above idea and consider having one variable for each edge in $H$, and a clause for each \emph{triangle} in $H$, where the three variables for each clause are given by the edges inside each triangle. We can again try 
to find a compressing subset by forcing all triangles to share as many edges as possible. Observe that, because any sufficiently small set of vertices has a near linear number of edges, we can only force all triangles to share a single edge. Nearly all of the other edges will be unique to their specific triangle. In this way, we get an upper bound on the expansion factor of $\gamma \leq \frac{2}{3}$. Extending this to having one variable for each $(k-1)$-clique and a clause of arity $k$ for each $k$-clique, the same worst-case construction gives an upper bound of $\gamma \leq \frac{k-1}{k}$. In fact, this this naive upper bound is close to the lower bound in the following sense. We show that for any constant $\varepsilon > 0$ there is a constant integer $k$ such that,
\begin{center}
    \textit{For any small enough set $I$ of clauses, nearly all clauses in $I$ have that a $(1-\varepsilon)$ fraction of their variables appear in at most $\polylog(n)$ many other clauses of $I$.}
\end{center} 

On the other hand, when $G$ contains a planted clique, we get a tiny subset $T$ of clauses where every variable is shared by a much higher number of clauses in $T$. Thus the planted clique gives a trapdoor ``highly compressing'' set of clauses, but when $G$ doesn't contain a planted clique, any small set of clauses can only be at worst mildly compressing. We refer readers to Lemma~\ref{lem:highdegree} for the formal claim and other details.

The above property takes us very close to our goal. 
However so far we cannot rule out the case that, for some small set of clauses $I$, the clauses together only touch $|I|/\polylog(n)$ many variables.

\paragraph{Augmentation with Random Vectors.} To ensure the stronger notion of expansion that we desire, we augment the clauses with random sparse vectors. In more detail, instead of each $(k-1)$-clique 
corresponding to just one variable, we will introduce a new constant parameter $\zeta$, and have each $(\zeta-1)$-clique correspond to $\log^C n$ many variables for some sufficiently large constant $C$. For every $\zeta$-clique, we will form a clause of arity $k = \Theta(\log |H|)$. To get the set of variables for each clause, we randomly subsample $k/\zeta$ variables from the set of $\log^C n$  variables corresponding to each $(\zeta-1)$-size sub-clique of the clause's $\zeta$-clique.

We show in Lemma~\ref{lem:randomexpand} that a set of random sparse vectors has exemplary small-set-expansion properties. Recall that, without augmentation, any small set of clauses is only mildly compressing -- nearly all of the clauses have a $(1-\varepsilon)$ fraction of their variables appearing in at most polylog$(n)$ other clauses. In Lemma \ref{thm:expanding}, we layer these two observations to show that augmentation gives the necessary expansion when $G$ is sampled from $\mathcal{G}(n, \frac{1}{2})$.

At the same time, when $G$ is sampled from $\cG(n,\frac{1}{2},n^\delta)$, our hidden ``highly compressing'' set of clauses is so dense that it remains dense even after this augmentation step, thereby enabling decryption (see proof of Lemma~\ref{lem:correct}).

\paragraph{Notational Guidance.} What we called $n$ here in this overview, we call $\bar{n}$ in the technical sections. We will use $n$ to refer to the total number of variables resulting from our final transformation as sketched above.

%% file: Prelims.tex
\section{Preliminaries}
\paragraph{Notations.} A vector is said to be $k$-sparse if it has exactly $k$ many non-zero entries. Similarly a matrix is called $k$-sparse if each of its rows  forms a $k$-sparse vector. Let $\hw(\mat v)$ be the Hamming weight of $\mat v$, so $\hw(\mat v) = k$ iff $\mat v$ is a $k$-sparse vector. Let $\bigvee_{\mat v \in S}\mat v$ denote the bitwise-OR of all vectors $\mat v \in S$. We use $[x]$ to denote the set $\{1, \ldots x\}$. 
We use $N_G(v)$ to denote the open neighborhood of vertex $v$ with respect to the graph $G$, i.e. all nodes (barring $v$) in $G$ that are connected to $v$ with an edge.
Throughout, the reader may assume that quantities are rounded to integer values as needed. All logarithms are taken base 2.

\begin{definition}[$(\gamma, k, t)$-Expanding Set System] \label{def:expand}
    Let $\mathcal{S}$ be a collection of sets. We say that $\mathcal{S}$ is a $(\gamma, k, t)$-expanding set system iff:
    \begin{enumerate}
        \item $\forall S \in \mathcal{S}, |S| = k$, and
        \item $\forall \mathcal{T} \subseteq \mathcal{S}$ such that $|\mathcal{T}| \leq t$: 
        \[\big|\cup_{S \in \mathcal{T}}S \big| \geq \gamma \cdot k|\mathcal{T}|.\] 
    \end{enumerate}
\end{definition}

\begin{definition}[$(\gamma, k, t)$-Expanding Matrix] 
    We say that $\mat S \in \{0, 1\}^{m \times n}$ corresponds to a $(\gamma, k, t)$-expanding set system iff:
    \begin{enumerate}
        \item For all rows $\mat S_i$ of $\mat S$, we have that $\hw(\mat S_i) = k$, and
        \item $\forall T \subseteq \{1, \ldots m\}$ such that $\vert T\vert \leq t$: 
        \[\hw\left(\bigvee_{i \in T}\mat S_i\right) \geq \gamma \cdot k\vert T\vert.\]
    \end{enumerate}
\end{definition}

\paragraph{Weak Public Key Encryption.} 
\begin{definition}[Adapted from Definition 4.1 in \cite{applebaum10}]
    An $(\psi(n), \phi(n))$-secure public-key bit encryption scheme is a triple $(\text{Gen}, \text{Enc}, \text{Dec})$ of probabilistic polynomial time algorithms such that
\begin{itemize}
    \item Algorithm $\text{Gen}$, on input $1^n$, produces a pair $(\text{pk}, \text{sk})$.
    
    \item $((1-\psi)$-correctness) For a random bit $b \leftarrow \{0,1\}$, 
    \[
    \Pr[\text{Dec}_{\text{sk}}(\text{Enc}_{\text{pk}}(b)) = b] > 1 - \psi(n),
    \]
    where $(\text{pk}, \text{sk}) \getsr  \text{Gen}(1^n)$ and the probability is over the randomness of $\text{Gen}$, $\text{Enc}$, $\text{Dec}$, and the choice of $b$.
    
    \item $(\phi$-privacy) The distributions $(\text{pk}, \text{Enc}_{\text{pk}}(0))$ and $(\text{pk}, \text{Enc}_{\text{pk}}(1))$ are $\phi(n)$-indistinguishable, where $(\text{pk}, \text{sk}) \getsr \text{Gen}(1^n)$.
\end{itemize}
If $\psi(n)$ and $\phi(n)$ 
satisfy $\psi < (1 - \sqrt{\phi})/2$ and $\phi$ is a constant, we say that the scheme is a \textit{weak PKE}.
\end{definition} 
\begin{lemma}[See Theorem 6 in \cite{C:HolRen05} ]
\label{lem:amp} Every weak PKE can be transformed into a semantically secure PKE that supports arbitrarily (polynomially) long messages.

\end{lemma}


\paragraph{Randomized Graph Product.} Our version of the randomized graph product is a slight variation of the original definition introduced in~\cite{berman92,manurangsi20}. This will be a crucial tool for the construction of the PKE. 


\begin{definition}[Randomized Graph Product]
    Given a graph $G = (V_G, E_G)$, integer $\ell$, and probability $p$, we define the randomized graph product $H \coloneqq (V_H, E_H) \getsr \RP_p^\ell(G)$ as follows:
    \begin{enumerate}
        \item Let $V' = (v_1,v_2,\ldots,v_{\binom{V_G}{\ell}})$  denote an ordered sequence of labels. Every label $v \in V'$ corresponds to a subset $U_{v} \subset V_G$ of size $\ell$ ordered in some fixed ``canonical" sense.
        \item Let $V_H$ be the node set of $H$ created in the following manner: for every $v \in V'$, add $v$ to $V_H$ independently with probability $p$.
        \item For each $u, v \in V_H$ with $u \neq v$, let $(u, v) \in E_H$ iff $U_{u} \cup U_{v}$ induces a clique in $G$. 
    \end{enumerate}
\end{definition}

%% file: assumptions.tex
\section{Our Hardness Conjectures}
\label{sec:ass}

In this section we redefine the conjectures and all of their variants formally as we use them in the paper.

\subsection{The Planted Clique Conjecture}

\begin{conjecture}[$n^{\log^\alpha n}$ hardness of the Planted Clique Conjecture]
There exist constants $0<\alpha<1$ and $\delta \in (0, \frac{1}{2})$ such that, for all $n^{O((\log n)^\alpha)}$-size algorithms $\cA$,

 \[\Bigg \vert \Pr[\cA(G)=1 \vert G \getsr \cG(n,\frac{1}{2})] - \Pr[\cA(G)=1 \vert G \getsr \cG(n,\frac{1}{2},n^\delta)] \Bigg \vert \leq \frac{1}{3}.\] 

\end{conjecture}

\begin{remark}
    Recall that, for conciseness, we simply refer to the above as ``the planted clique conjecture.''
\end{remark}

As discussed earlier, this conjecture is based on a long line of work in the average-case complexity community~\cite{alon98,jerrum92,kuvcera1995expected,karp76,ITCS:ERSY22,juels2000hiding,applebaum10,TCC:ABIKN23,manurangsi20,hazan2011hard,STOC:AAKMRX07,arora2011computational,berthet2013complexity,brennan2018reducibility,brennan2020reducibility,STOC:HirShi24,ma2015computational,koiran2014hidden,hajek2015computational}.

\subsection{Noisy $k$-$\mathsf{LIN}$ Conjecture}
We first restate the standard noisy $k\mbox{-}\mathsf{LIN}$ conjecture from the introduction. Recall that this conjecture is substantiated by a long series of works giving lower bounds on the $k\mbox{-}\mathsf{LIN}$ problem~\cite{grigoriev01,FOCS:FeiKimOfe06,schoenebeck08,barak2016noisy,STOC:RagRaoSch17,STOC:FelPerVem15,applebaum2016cryptographic,bogdanov2019xor,STOC:KMOW17}.

\begin{conjecture}[Noisy $k$-$\mathsf{LIN}$ Conjecture] \label{con:dwdexpanderkosher}
There exists a constant $\gamma \in (\frac{1}{2}, 1)$ such that the following holds.
Let $\alpha, \varsigma$ be any constants with $\alpha \in (0,1)$ and $\varsigma>0$.
Then, for $n$ large enough, prime $p = n^{\Omega(1)}$, $k=\Omega(\log(n))$, we have the following: For all $(\gamma, k, 2^{(\log n)^\alpha})$-expanding matrices $\overline{\mat S} \in \{0, 1\}^{m \times n}$ and for all polynomial sized adversaries $\cA$, \[\left \vert \Pr_{\mathbf{A}, \mathbf{s}, \mathbf{e}}[\cA(\mat A, \mat b)=1\mid \mat s \getsr \bF_p^n, \mat b=\mat A\mat s+\mat e] - \Pr_{\mathbf{A}, \mathbf{b}}[\cA(\mat A, \mat b)=1\mid  \mat b\getsr \bF_p^m] \right \vert \leq \frac{1}{4}.\]
In the above, $\mathbf{A} \in \mathbb{F}_p^{m \times n}$ is the matrix produce by replacing each nonzero entry in $\overline{\mathbf{S}}$ with a random element in $\mathbb{F}_p^*$, and $\mathbf{e}$ is sampled as:
\[\mathbf{e}_i=\begin{cases} \text{Random element of $\mathbb{F}_{p}^*$, with probability $\eta= (\log n)^{-\varsigma}$. } \\ 0, \text{ otherwise.} \end{cases}\]
\end{conjecture}

Below we define the analogous search version of the noisy $k\mbox{-}\mathsf{XOR}$ conjecture which we use to construct another PKE in Section~\ref{sec:search}.

\begin{conjecture}[Search Noisy $k\mbox{-}\mathsf{LIN}$ Conjecture]\label{con:sexpander}

There exists a constant $\gamma \in (\frac{1}{2}, 1)$ such that the following holds.
Let $\alpha, \varsigma$ be any constants with $\alpha \in (0,1)$ and $\varsigma>0$.
Then, for $n$ large enough, prime $p = n^{\Omega(1)}$, $k=\Omega(\log(n))$, we have the following: For all $(\gamma, k, 2^{(\log n)^\alpha})$-expanding matrices $\overline{\mat A} \in \{0, 1\}^{m \times n}$ and for all polynomial sized adversaries $\cA$, \[\Pr_{\mathbf{A}, \mathbf{s}, \mathbf{e}}[\cA(\mat A, \mat b)=\mat s\mid \mat s \getsr \bF_p^n, \mat b=\mat A\mat s+\mat e] \leq \frac{1}{4}.\]
In the above, $\mathbf{A} \in \mathbb{F}_p^{m \times n}$ is the matrix produce by replacing each nonzero entry in $\overline{\mathbf{A}}$ with a random element in $\mathbb{F}_p^*$, and $\mathbf{e}$ is sampled as:
\[\mathbf{e}_i=\begin{cases} \text{Random element of $\mathbb{F}_{p}^*$, with probability $\eta= (\log n)^{-\varsigma}$. } \\ 0, \text{ otherwise.} \end{cases}\]

\end{conjecture}

In other words, it should be hard to recover $\mat s$ given $(\mat{A}$, $\mat{A}\mat{s}+\mat{e})$. In Section~\ref{sec:search} we present a PKE construction from Conjecture~\ref{con:plantedclique} and Conjecture~\ref{con:sexpander}. While a reduction from search to decision noisy $k\mbox{-}\mathsf{LIN}$ exists for random expander matrices $\mat{A}$~\cite{applebaum2016cryptographic,bogdanov2019xor}, the reduction suffers from a polynomial loss in the sample complexity, i.e. the number of clauses. Thus our construction in Section~\ref{sec:search} relies on a potentially weaker\footnote{The assumption is weaker in the sense that the search noisy $k$-$\mathsf{LIN}$ is at least as hard as decision noisy $k$-$\mathsf{LIN}$.} assumption that the one in Section~\ref{sec:pke} for the following two reasons: (1) the semantic security reduction to the search version does not suffer from a polynomial loss, and (2) it is unclear if the search-decision reduction mentioned above extends to arbitrary expander matrices $\mat{A}$.

\paragraph{Non-Compression.} In fact, the matrix $\mat A$ from both of our PKE schemes satisfies an additional property: The matrix $\mat S \in \{0, 1\}^{m \times n}$ indicating the nonzero entries in $\mathbf{A}$ is a $(\gamma, \Theta(\log n))$-non-compressing matrix, for any constant $0 < \gamma < 1$ fixed at the beginning of key generation. We define this property below.

\begin{definition}[$(\gamma, k)$-Non-Compressing Matrix] \label{def:compress}
    We say that $\mat S \in \mathbb{F}_2^{m \times n}$ is a $(\gamma, k)$-non-compressing matrix iff:
    \begin{enumerate}
        \item For all rows $\mat S_i$ of $\mat S$, we have that $\hw(\mat S_i) = k$, and
        \item $\forall T \subseteq \{1, \ldots m\}$: 
        \[\hw\left(\bigvee_{i \in T}\mat S_i \right)\geq |T|^{\gamma}.\]
    \end{enumerate}
\end{definition}

In other words, for \emph{every arbitrarily large} set of rows, the sum of hamming weights for those rows should not be much greater than the hamming weight when we bitwise-OR all of the rows together. This property is desirable in the context of Goldreich's candidate one-way function \cite{goldreich11, applebaum12, applebaum16}, which is highly related to noisy $k\mbox{-}\mathsf{LIN}$. (In this context ``non-compression'' is known as ``low stretch.'')

%% file: PKE.tex
\section{Public key encryption from Planted Clique and Noisy $k\mbox{-}\mathsf{LIN}$}
\label{sec:pke}
In this section, we give a PKE scheme from the \emph{decisional} noisy $k\mbox{-}\mathsf{LIN}$ conjecture. For a description of how to use the search version, see Section \ref{sec:search}.

\paragraph{Suggested Parameters.} We suggest the following choices of parameters. The constants are solely optimized for exposition, and we believe they can be improved significantly while still maintaining the integrity of the scheme. Let $\alpha, \delta \in (0, \frac{1}{2})$ be constants such that Conjecture \ref{con:plantedclique} holds, and let $\gamma \in (\frac{1}{2}, 1)$ be a constant such that Conjecture \ref{con:dwdexpanderkosher} holds. (If Conjecture \ref{con:plantedclique} holds for some $\alpha \geq \frac{1}{2}$, this automatically implies that the conjecture holds for all $\alpha \in (0, \frac{1}{2})$.) Assuming the security parameter is $\lambda$, we will set the following parameters: 

\begin{itemize}
    \item $p$ is any prime.
    \item Size of initial Erdos-Renyi random graph: $\bar{n} = 2^{(\log\lambda)^{1/(1+\alpha)}}$.
    \item Planted clique size: $\bar{k} = \bar{n}^\delta$.
    \item Graph product parameter for $\RP$: $\ell=(\log\bar{n})^\alpha$.
    \item Node deletion probability for $\RP$: $q=\bar{n}^{-(\delta\ell)/2}$.
    \item Clique size (constant integer) for constructing $k\mbox{-}\mathsf{LIN}$ matrix: $\zeta = \lceil\frac{1}{1-\sqrt{\gamma}}\rceil + 1$.
    \item Exponent (constant) for block size in $k\mbox{-}\mathsf{LIN}$ matrix: $\beta = \zeta \cdot (\zeta(1-\alpha)+2)$.
    \item Exponent (constant) for Bernoulli noise for $k\mbox{-}\mathsf{LIN}$ samples: $\rho=2+\beta\zeta$.
\end{itemize}

\begin{remark}
    The final three parameters are balanced so that, if our final $k\mbox{-}\mathsf{LIN}$ matrix $\mat A \in \mathbb{F}_2^{m \times n}$ is constructed from an Erdős–Rényi graph without a planted clique, we achieve arbitrarily high expansion on all sets of at most $t = 2^{(\log n)^\alpha}$ many rows, while maintaining an error rate of $(\log t)^{-O(1)}$. At the same time, we achieve arbitrarily high non-compression over the entire matrix.
\end{remark}

\paragraph{Notations.} Let $H \getsr \RP^\ell_q(G)$. Then, by construction, every node $v \in V_H$ corresponds to a set $U_{v} \subset V_G$ of size $\ell$ in some ``canonical" order. Let $c_{H,\zeta}$ denote the total number of $\zeta$-cliques in a graph $H$. Let $\mat{A}_{i}$ denote the $i^{th}$ row of a matrix $\mat{A}$. Allow $\|$ to denote the concatenation of two vectors in the standard sense. We abuse notation and allow $(G,K_G) \getsr \cG(\bar{n},\frac{1}{2},\bar{k})$ to denote a graph $G$ drawn from the planted Erdős–Rényi distribution along with the set of vertices $K_G$ for the planted clique.

   

\paragraph{Encoding Process.} First we describe a randomized encoding of any graph $H$ with $n$ nodes into a sparse matrix $\mat{S} \in \{0, 1\}^{m \times n}$, where $m = c_{H,\zeta}$ and $n = c_{H,\zeta-1}\cdot \log^\beta m$. We use this algorithm in the key generation step of the PKE.

\begin{mdframed}
\begin{itemize}
     \item[$\star$] $\mat{A} \from \Encode(H,\zeta,\beta):$
     \begin{enumerate}
         \item Let $m = c_{H,\zeta}$. Enumerate over all $\zeta$-cliques of $H$ in a fixed canonical order and create labels $(R_1,R_2,\ldots,R_{m})$.
         \item Let $n' = c_{H,\zeta-1}$. Enumerate over all $(\zeta-1)$-cliques of $H$ in a fixed canonical order and create labels $(C_1,C_2,\ldots,C_{n'})$.
         \item For all $i \in [m], j \in [n']$ do the following:
         \begin{itemize}
             \item Set an indicator variable $I_{i,j}=1$ if the $(\zeta-1)$-clique labeled $C_j$ is a subgraph of the $\zeta$-clique labeled $R_i$, and set $I_{i,j}=0$ otherwise.
         \end{itemize}
         \item For all $i \in [m], j \in [n']$ do the following:
         \begin{itemize}
             \item Set $\mat{b}_{i,j} \in \{0, 1\}^{(\log m)^\beta}$ to be a random $\frac{\log m}{\zeta}$-sparse vector if $I_{i, j} = 1$, and set $\mat{b}_{ij} = \mat{0}^{\log ^\beta m}$ otherwise. We refer to each $\mat b_{i, j}$ as a \emph{block}.
         \end{itemize}
         \item For all $i \in [m]$, set $\mathbf{S}_{i} \coloneqq (\mat{b}_{i,1}\|\mat{b}_{i,2}\|\cdots\|\mat{b}_{i,n'})$.
         \item Output $\mathbf{S}$.
     \end{enumerate}
\end{itemize}
\end{mdframed}

\paragraph{Key Generation.}
\begin{mdframed}
\begin{itemize}
    \item[$\star$] $(\pk, \sk)\from \Gen(1^{\lambda})$:
    \begin{enumerate}
        \item $(G,K_G) \getsr \cG(\bar{n},\frac{1}{2},\bar{k})$.
        \item $H \getsr \RP^\ell_q(G)$.
        \item Define $K_H=\{v \in V_H: U_{v} \subset K_G\}$. 
        \item $\overline{\mathbf{A}} \gets \Encode(H,\zeta,\beta)$. Let $\mathbf{S}$ have $m$ rows and $n$ columns.
        Let $\mathbf{A} \in \mathbb{F}_p^{m \times n}$ be the matrix where we replace each nonzero entry in $\overline{\mathbf{A}}$ with a random element in $\mathbb{F}_p^*$.
        \item $\pk \coloneqq \mat{A}$.
        \item $\sk \coloneqq (H, K_H)$.
        \item Output $(\pk, \sk)$.
    \end{enumerate}
\end{itemize}
\end{mdframed}

\paragraph{Encryption.}
\begin{mdframed}
\begin{itemize}
    \item[$\star$] $\ct \from \Enc(1^\secParam, \pk, b \in \{0, 1\})$:
    \begin{enumerate}
        \item Parse $\pk$ as $\mat{A}$. Let $\mat{A}$ have $m$ rows and $n$ columns. 
        \item Sample $\mat{e}$ as:
        \[\mathbf{e}_i=\begin{cases} \text{Random element of $\mathbb{F}_{p}^*$, with probability $\eta= (\log n)^{-\varsigma}$. } \\ 0, \text{ otherwise.} \end{cases}\]
        \item Sample $\mat{s} \getsr \bF_p^{n}$.
        \item If $b=0$, then $\ct \gets \mat{A}\mat{s}+\mat{e}$.
        \item If $b=1$, then $\ct \getsr \bF_2^{m}$.
        \item Output $\ct$.
    \end{enumerate}
\end{itemize}
\end{mdframed}

\paragraph{Decryption.}
\begin{mdframed}
\begin{itemize}
    \item[$\star$] $b' \from \Dec(1^\secParam, \pk, \sk, \ct)$:
    \begin{enumerate}
        \item Parse $\pk$ as $\mat{A}$. Let $\mat{A}$ have $m$ rows and $n$ columns.
        \item Parse $\sk$ as $(H, K_H)$.
        \item Parse $\ct$ as $\mat b$.
        \item Let $\eta$ be any (superconstant) integer satisfying $(\log n)^{\rho-1} \leq \binom{\eta}{\zeta} \leq 2(\log n)^{\rho-1}$.
        \item Uniformly randomly select a subset $T \subset K_H$ among all subsets of size $\eta$.
        \item Recompute all canonical labels $R_i$ and $C_j$. Recompute the indicator variables $I_{i, j}$.
        \item Define $X^{(\zeta)}_i$ as the subset of $V_H$ corresponding to the $\zeta$-clique labeled as $R_i$. Define a set $S$ of row labels of $\mat{A}$ as \[S=\left\{i: X^{(\zeta)}_i \subset T\right\}\] 
        \item Split every row of $\mat{A}$ into blocks of length $\log^\beta m$, i.e., $\mat{A}_{i}=(\mat{b}_{i,1}\|\cdots\|\mat{b}_{i,c_{H,\zeta-1}})$ where every $\mat{b}_{i,j}$ has width $\log^\beta m$. Define a set of block labels $S' \subset [c_{H,\zeta-1}]$ such that \[S'=\{j: \exists i \in S \text{ such that } I_{i,j} = 1\}.\]
        \item Let $\tilde{\mat{A}}$ be the submatrix of $\mat{A}$ with rows labeled by $S$ and columns made by concatenating the blocks labeled by $S'$.
        \item Solve the following system of linear equations using Gaussian elimination:\[\tilde{\mat{A}}\mat{x}=\mat b\vert_S,\] where $\mat b\vert_S$ is the vector $\mat b$ restricted to indices in $S$.
        \item If the above step returns a valid solution, then output $b'=0$. Otherwise output $b'=1$. 
    \end{enumerate}
\end{itemize}
\end{mdframed}

\subsection{Correctness}
First, we show that $K_H$ is almost surely very large.
\begin{lemma}\label{lem:large}
    For all constants $\alpha, \delta \in (0, \frac{1}{2})$, for all large enough $\lambda$, the following holds. Let $\bar{n} = 2^{(\log\lambda)^{1/(1+\alpha)}}$, $\bar{k} = \bar{n}^\delta$, $\ell=(\log\bar{n})^\alpha$, and $q=\bar{n}^{-(\delta\ell)/2}$.
    Then the set $K_H$ produced in algorithm $\Gen$ satisfies \[\vert K_H\vert > |V_H|^{\delta/2}\] with all but $o(1)$ probability over the randomness of $\RP_p^\ell$.
\end{lemma}
\begin{proof}
    The planted clique contains $\binom{\bar{k}}{\ell}$ many $\ell$-tuples of vertices, and the entire graph $G$ contains $\left(^{\bar{n}}_\ell\right)$ many $\ell$-tuples. Each $\ell$-tuple becomes a vertex in $H$ independently with probability $\bar{n}^{-\delta\ell/2}$, so a Chernoff argument shows that $|V_H| = \bar{n}^{(1-\delta/2 \pm o(1))\ell}$ with all but $o(1)$ probability. Since $\bar{k} = \bar{n}^{\delta}$, another Chernoff argument shows that the number of $\ell$-tuples inherited from the planted clique is $\bar{n}^{(\delta/2 \pm o(1))\ell} = |V_H|^{\delta/(2-\delta) \pm o(1)} > |V_H|^{\delta/2}$ with all but $o(1)$ probability, with the last inequality holding for all large enough $\lambda$.
\end{proof}

We use the above to argue decryption correctness.

\begin{lemma}\label{lem:correct}
For all constants $\alpha, \delta \in (0, \frac{1}{2})$, for all large enough $\lambda$, the following holds. Let $\bar{n} = 2^{(\log\lambda)^{1/(1+\alpha)}}$, $\bar{k} = \bar{n}^\delta$, $\ell=(\log\bar{n})^\alpha$, $q=\bar{n}^{-(\delta\ell)/2}$, $\zeta > 1$ be a constant integer, $\beta = \Theta(1)$, and $\rho=2+\beta\zeta$. Then for all $b \in \{0, 1\}$
\[\Pr_{(\pk, \sk) \from \Gen(1^{\bar{n}})} [\Dec(\sk, \Enc(\pk, b)) = b) ] \geq 1-o(1),\] where the probability is over the coins of $\Encode$, $\Gen$, $\Enc$, and $\Dec$.
\end{lemma}
\begin{proof}
    Let $\mat A$ have $m$ rows and $n$ columns. First observe that $K_H$ is made of nodes that form a subgraph of $K$ in $G$, and therefore $K_H$ must form a clique. Moreover by Lemma~\ref{lem:large}, the size of $K_H$ is more than $|V_H|^{\delta/2}$ with all but $o(1)$ probability.\\\\
    Since $\rho$ is a constant, we can always find an integer $\eta$ satisfying the requirement for decryption. By construction, $|T| = \eta$, and $T$ induces a clique in $H$. Moreover, $S$ contains precisely all of the row labels for $\zeta$-cliques that are subgraphs of $T$, and $\vert S \vert=\binom{\eta}{\zeta}$.\\\\
    The encoding of $H$ to $\mat{A}$ is such that for any row $\mat{A}_{i}$, the $j^{th}$ block $\mat{b}_{i,j}$ is non-zero (i.e. $I_{i,j}=1$) iff the $(\zeta-1)$-clique labeled as $C_j$ is a subgraph of the $\zeta$-clique labeled as $R_i$. As $S$ contains all $\zeta$ sized cliques that are a subgraph of $T$, and $T$ is a clique, the set $S$ corresponds to row labels of all possible $\zeta$-cliques that are subgraphs of $T$. This count is exactly $\binom{\eta}{\zeta}$. A similar argument shows that the size of $S'$ is exactly $\binom{\eta}{\zeta-1}$.
    Thus, the total number of columns in $\tilde{\mat{A}}$ is $\binom{\eta}{\zeta-1} \cdot \log ^\beta m$, and the total number of rows is $\binom{\eta}{\zeta}$. \\\\
    Note that for every choice of $\mat{x} \in \bF_p^{\binom{\eta}{\zeta-1} \cdot \log ^\beta m}$, we have that $\tilde{\mat{A}}\mat{x}$ can map to at most one unique vector in $\bF_p^{\binom{\eta}{\zeta}}$. Also, since $\zeta$ is a constant and $\binom{\eta}{\zeta} = \Theta((\log n)^{\rho-1})$, we have that $\eta = \Theta((\log n)^{(\rho-1)/\zeta})$ by the Stirling approximation.
    Thus if we encrypted a $1$, i.e. $\mat{b}\vert_S$ is a random vector in $\bF_p^{\vert S \vert}$, then the probability that Gaussian elimination will return a solution is upper bounded as follows.
    
    \begin{align*}
        \Pr[\text{returns solution}] \leq & \frac{p^{\binom{\eta}{\zeta-1} \cdot \log ^\beta m}}{p^{\binom{\eta}{\zeta}}} \\
        \leq & \frac{p^{\binom{\eta}{\zeta-1} \cdot \log ^\beta m}}{p^{\binom{\eta}{\zeta}}} \\
        \leq & 2^{\binom{\eta}{\zeta-1} \cdot \log ^\beta m - \binom{\eta}{\zeta}} \\
        < & 2^{\eta^{\zeta-1} \cdot \log ^\beta m - \Omega(\eta^{\zeta})}  \tag{Stirling approximation}\\
        \leq & 2^{O((\log n)^{(\rho-1) \cdot (\zeta-1)/\zeta} \cdot \log ^\beta m) - \Omega((\log n)^{\rho-1})} \\
        < & 2^{O((\log n)^{(\rho-1) \cdot (\zeta-1)/\zeta} \cdot (\log n)^\beta) - \Omega((\log n)^{\rho-1})} \tag{$m$ and $n$ are polynomially related w.h.p., and $\beta$ is constant}\\
        \leq & 2^{O((\log n)^{(1 + \beta\zeta) \cdot (\zeta-1)/\zeta + \beta}) - \Omega((\log n)^{1 + \beta\zeta})} \tag{$\rho = 2 + \beta\zeta$}\\
        \leq & 2^{O((\log n)^{1 + \beta\zeta - 1/\zeta}) - \Omega((\log n)^{1 + \beta\zeta})} \\
        \leq & 2^{- \Omega((\log n)^{1 + \beta\zeta})}  \tag{$\zeta$ is a constant}\\
    \end{align*}
    With (very) high probability, $n$ is superconstant in $\lambda$. Therefore, \[\Pr [(\pk, \sk) \from \Gen(1^n) \land \Dec(\sk, \Enc(\pk, 1)) = 1) ] \geq 1-o(1).\] \\
    
    Let us look at the case that a $0$ was encrypted. Here $\mat{b}=\mat{A}\mat{s}+\mat{e}$. Let $S_{\text{col}}$ be the set of columns for the blocks in $S'$. The linear system solved by Gaussian elimination becomes \[\tilde{\mat{A}}\mat{x}=\tilde{\mat{A}} \cdot (\mat{s}\vert_{S_{\text{col}}})+\mat{e}\vert_S.\] Now, every entry of $\mat{e}$ has is nonzero independently with probability at most $(\log n)^{-\rho}$. So the probability that $\mat{e}\vert_S=\mat{0}$ is at most \[(1-(\log n)^{-\rho})^{\vert S \vert} = (1-(\log n)^{-\rho})^{\binom{\eta}{\zeta}} \geq (1-(\log n)^{-\rho})^{\Theta((\log n)^{(\rho-1)})} \geq 1 -o(1),\] again using that $n$ is superconstant in $\lambda$ with (very) high probability.\\\\
    If this happens, then Gaussian elimination must return a valid solution. If there is at least one error, it is still possible that we get a valid solution, but this only increases the decryption correctness. Therefore \[\Pr [(\pk, \sk) \from \Gen(1^n) \land \Dec(\sk, \Enc(\pk, 0)) = 0) ] \geq 1-o(1).\] We conclude that for all $b\in \{0,1\}$, \[\Pr [(\pk, \sk) \from \Gen(1^n) \land \Dec(\sk, \Enc(\pk, b)) = b) ] \geq 1-o(1).\] 
\end{proof}




\subsection{Security}
The proof of security will follow from standard hybrid arguments by reducing to the hardness of planted clique (Conjecture~\ref{con:plantedclique}) and noisy $k\mbox{-}\mathsf{LIN}$ (Conjecture~\ref{con:dwdexpanderkosher}). To use the latter, we first need to show that the indicator matrix $\mathbf{S}$ for $\mat{A}$ indeed satisfies the pre-conditions of Conjecture~\ref{con:dwdexpanderkosher}. Namely, $\mathbf{S}$ should be an expanding matrix when $\mathbf{S}$ is constructed from a graph $G$ without a planted clique. We will also show that $\mathbf{S}$ is a non-compressing matrix when constructed in this manner. We present the proofs below.

\subsubsection{Proof of Expansion}

Before showing the proof of expansion, we prove the following lemma regarding sparse vectors.

\begin{lemma} \label{lem:randomexpand}
    Let $c_1, c_2 > 0$ be any constants. Let $n \in \bN$ and $\Delta = (\log n)^{c_2}$. Let $T$ be a set of $n^{O(1)}$ many uniformly sampled $t$-sparse vectors in $\bF_2^{(\log n)^{c_1}}$, where $t = \Theta(\log n)$. Then, for all subsets $S \subset T$ of size at most $\Delta$, \[\hw(\bigvee_{\mat s \in S} \mat s) > \frac{c_1-c_2-2}{c_1} \cdot \sum_{\mat s \in S}\hw(\mat s)\] with all but $o(1)$ probability over the choice of the vectors.
\end{lemma}

\begin{proof}
    Clearly the lemma holds if $c_1-c_2-2 \leq 0$. So consider the case that $c_1-c_2-2 > 0$.\\\\
    We use linearity of expectation to calculate the number of sets of vectors $S$ with $|S| \leq \Delta$ such that \[\hw(\bigvee_{\mat s \in S} \mat s) \leq \frac{c_1-c_2-2}{c_1} \cdot \sum_{\mat s \in S}\hw(\mat s) = \frac{c_1-c_2-2}{c_1} t |S|.\] Let $x := |S|$ and the number of sets of size $x$ that contradict the lemma be $N_x$. We want all non-zero entries of each $\mat s \in S$ to be restricted to a subset of $[(\log n)^{c_1}]$ that has size at most $\frac{c_1-c_2-2}{c_1} t x$. \\\\
    Now, the total number of ways to choose a set of $x$ many vectors is $\binom{n^{O(1)}}{x} \leq n^{O(x)}$, and the total number of ways to select $\frac{c_1-c_2-2}{c_1} t x$ many positions out of a total of $(\log n)^{c_1}$ positions is \[\binom{(\log n)^{c_1}}{\frac{c_1-c_2-2}{c_1} t x} < ((\log n)^{c_1})^{\frac{c_1-c_2-2}{c_1} t x}.\]
    The probability that all the non-zero entries for all vectors $\mat x \in S$ lie within a given set of $\frac{c_1-c_2-2}{c_1} t x$ many positions is \[\left(\frac{\binom{\frac{c_1-c_2-2}{c_1}tx}{t}}{\binom{(\log n)^{c_1}}{t}}\right)^x < \left(\frac{\frac{c_1-c_2-2}{c_1} t x}{(\log n)^{c_1}}\right)^{t x} < \left(\frac{t x}{(\log n)^{c_1}}\right)^{t x} < \left(\frac{(\log n)^{3/2}\Delta}{(\log n)^{c_1}}\right)^{t x} = ((\log n)^{3/2+c_2-c_1})^{t x}.\]
    Putting all of this together, we have:
    \[\begin{split}
        \mathbb{E}[N_x] \leq & n^{O(x)} \cdot ((\log n)^{c_1})^{\frac{c_1-c_2-2}{c_1} t x} \cdot ((\log n)^{3/2+c_2-c_1})^{t x} \\
        \leq & 2^{O(x\log n)} \cdot 2^{(c_1-c_2-2) t x \log\log n} \cdot 2^{(3/2+c_2-c_1) t x \log\log n} \\
        \leq & 2^{O(x\log n)} \cdot 2^{((c_1-c_2-2) + (3/2+c_2-c_1)) t x \log\log n} \\
        \leq & 2^{O(x\log n)} \cdot 2^{-\frac{1}{2} t x \log\log n} \\
        \leq & 2^{O(x\log n)} \cdot 2^{-\omega(x \log n)} \\
        \leq & 2^{-\omega(x \log n)} \\
    \end{split}\]
    Finally taking a union bound of all $x \leq \Delta$ and applying Markov's inequality, we get that
    \[\Pr[N\coloneqq\sum_xN_x \geq 1] \leq 2^{-\omega(\log n)}.\]

\end{proof}

Next, we state a useful lemma about $\zeta$-cliques and $(\zeta-1)$-cliques, the proof of which is deferred to Section \ref{sec:randomgraphs}. To state the lemma, we begin by defining a notion of ``high'' and ``low'' degree that applies to cliques.

\paragraph{Definition \ref{def:highdegree} \textnormal{($(\Delta, S)$-High Degree)}.} \emph{A $(\zeta-1)$-clique is said to be \emph{$(\Delta, S)$-high degree} if it is contained in at least $\Delta$ many distinct $\zeta$-cliques of $S$, where $S$ is a set of $\zeta$-cliques.}

\paragraph{Lemma \ref{lem:highdegree}.} \emph{Let $\bar{n} \in \mathbb{N}$ be any parameter, $\zeta > 1$ be any (not necessarily constant) integer, and $\alpha, \delta \in (0, \frac{1}{2})$ be any constants. Set $\ell=(\log\bar{n})^\alpha$, $q=\bar{n}^{-(\delta\ell)/2}$, and $\Delta = (\log\bar{n})^{\zeta(1-\alpha)}$. Sample $G \getsr \mathcal{G}(\bar{n}, \frac{1}{2})$, and sample $H \getsr \RP_q^\ell(G)$. With all but $o(1)$ probability over the random choice of $G$ and randomness of $\RP_q^\ell$, the following occurs. For any set $S$ of $2^{o((\log\bar{n})^{2\alpha})}$ many $\zeta$-cliques of $H$, define $T \subset S$ as \[T=\{K \in S: K \text{ has as a subgraph more than one $(\Delta, S)$-high degree $(\zeta-1)$-clique}\}.\] Then $\vert T \vert=o(1) \cdot \vert S \vert$.}\\\\
In other words, the above lemma says that for any set $S$ consisting of $\zeta$-cliques, as long as the size of $S$ is not too high, nearly all of the $\zeta$-cliques in $S$ contain at most one high degree $(\zeta-1)$-clique.\\\\
Using the above lemma, we can argue that $\mat A$ is a good small-set-expander.

\begin{lemma}\label{thm:expanding}
    For all constants $\alpha, \delta \in (0, \frac{1}{2})$ and $\gamma \in (\frac{1}{2}, 1)$, for all large enough $\lambda$, the following holds. Set $\bar{n} = 2^{(\log\lambda)^{1/(1+\alpha)}}$, $\ell=(\log\bar{n})^\alpha$, $q=\bar{n}^{-(\delta\ell)/2}$, $\zeta = \lceil\frac{1}{1-\sqrt{\gamma}}\rceil + 1$, and $\beta = \zeta \cdot (\zeta(1-\alpha)+2)$. Let $G \getsr \mathcal{G}(\bar{n}, \frac{1}{2})$, $H \getsr \RP_q^\ell(G)$, and $\mat{A} \gets \Encode(H,\zeta,\beta)$. Then $\mat{A} \in \mathbb{F}_2^{m \times n}$ is a $(\gamma, \Theta(\log n), 2^{(\log n)^\alpha})$-expanding matrix with $1 - o(1)$ probability over the random choice of $G$, the randomness of $\RP^\ell_q$, and the coins of $\Encode$.
\end{lemma}

\begin{proof}
First note that, by construction, each row of $\mat A$ is exactly $k$-sparse, where $k = \log m = \Theta(\log n)$. ($m$ and $n$ are polynomially related with probability $1 - o(1)$ over the random choice of $G$ and the randomness of $\RP_p^\ell$.)\\\\
Recall that our small-set-expansion requirement is the following, where $t = 2^{(\log n)^\alpha}$.
\[\forall S \subseteq \{1, \ldots m\} \text{ such that } |S| \leq t, \text{ we have that } \hw(\vee_{i \in S}\mathbf{S}_i) \geq \gamma \cdot k|S|.\]
Consider any set of indices $S$ with $|S| \leq t$. Define $B$ to be the set of all blocks $\mat{b}_{i, j}$ such that $i \in S$ and $I_{i, j} = 1$. We now lower bound $\hw(\vee_{i \in S}\mathbf{S}_i)$ by analyzing the blocks in $B$. \\\\
Fix $\Delta = (\log \bar{n})^{\zeta(1-\alpha)}$. Let $K(S) = \{\zeta$-clique of $H$ with label $R_i$ $: i \in S\}$. Let $S_{\text{Bad}} \subset S$ be the set of all $i \in S$ such that the $\zeta$-clique labeled $R_i$ contains at least two $(\Delta, K(S))$-high degree $(\zeta-1)$-cliques.\\\\
Observe that, since $n = |V_H|^{O(1)} = \bar{n}^{O(\ell)}$ and $\alpha \in (0, \frac{1}{2})$ is a constant, we have that $\log ^\alpha n = o(\log ^{2\alpha} \bar{n})$. So $|K(S)| \leq 2^{o(\log ^{2\alpha} \bar{n})}$. Therefore by Lemma \ref{lem:highdegree}, 
\[|S_{\text{Bad}}| \leq o(1) \cdot |S|\]
Define $B_{\text{Good}}$ to be the set of all blocks $\mat{b} \in B$ that are contained in a row $\mathbf{S}_i$ with $i \in S\backslash S_{\text{Bad}}$. Since every row $\mathbf{S}_i$ contains exactly $\zeta$ many blocks $\mat{b}_{i, j}$ with $I_{i, j} = 1$:
\[|B_{\text{Good}}| \geq (1-o(1)) \cdot |B|.\]
Let $B_{\text{Good, Low}}$ be the set of all blocks $\mat{b}_{i, j} \in B_{\text{Good}}$ such that $C_j$ is the label of a $(\zeta-1)$-clique that is \emph{not} $(\Delta, K(S))$-high degree. Since each $\zeta$-clique with label $R_i$ for some $i \in S\backslash S_{\text{Bad}}$ has that all but one of its $(\zeta-1)$-cliques are not $(\Delta, K(S))$-high degree,
\[|B_{\text{Good, Low}}| \geq \frac{\zeta-1}{\zeta} \cdot |B_{\text{Good}}| \geq \frac{\zeta-1}{\zeta} \cdot (1-o(1)) \cdot |B|.\]
Since each block has exactly the same hamming weight, it must be the case that
\[\sum_{\mat b \in B_{\text{Good, Low}}} \hw(\mat b) = \frac{|B_{\text{Good, Low}}|}{|B|} \cdot \sum_{\mat b \in B} \hw(\mat b).\]
Furthermore
\[\sum_{\mat b \in B} \hw(\mat b) = \sum_{i \in S} \hw(\mat A_i) = k|S|\]
since all of the nonzero entries in each row appear in a block $\mat b_{i, j}$ with $I_{i, j} = 1$. Composing these results, we have
\[\sum_{\mat b \in B_{\text{Good, Low}}} \hw(\mat b) \geq \frac{\zeta-1}{\zeta} \cdot (1-o(1)) \cdot k|S|.\]\\

Let $B_{\text{Good, Low}, j}$ be the set of all $\mat{b}_{i, j} \in B_{\text{Good, Low}}$ with a fixed $j$ value. Recall that each block in $B_{\text{Good, Low}, j}$ is a $(\log m)$-sparse vector in $\mathbb{F}_2^{\log^\beta m}$. Furthermore, since $m > \bar{n}$ w.h.p, $B_{\text{Good, Low}, j}$ contains at most $\Delta = (\log \bar{n})^{\zeta(1-\alpha)} < (\log m)^{\zeta(1-\alpha)}$ many vectors selected out of a total of $m^{O(1)}$ many vectors. (The set of vectors is taken to be the set of all blocks $\mat b_{i, j}$ in $\mathbf{S}$.) We now apply Lemma \ref{lem:randomexpand} with $c_1 = \beta = \zeta \cdot (\zeta(1-\alpha)+2)$ and $c_2 = \zeta(1-\alpha)$. For each $j$, we have that
\[\begin{split}
    \hw(\bigvee_{\mat b \in B_{\text{Good, Low}, j}}{\mat b}) > & \frac{\zeta \cdot (\zeta(1-\alpha)+2) - \zeta(1-\alpha) - 2}{\zeta \cdot (\zeta(1-\alpha)+2)} \cdot \sum_{\mat b \in B_{\text{Good, Low}, j}}\hw(\mat b) \\
    = & \frac{\zeta \cdot (\zeta(1-\alpha)+2) - 1 \cdot (\zeta(1-\alpha) + 2)}{\zeta \cdot (\zeta(1-\alpha)+2)} \cdot \sum_{\mat b \in B_{\text{Good, Low}, j}}\hw(\mat b) \\
    = & \frac{(\zeta -1) \cdot (\zeta(1-\alpha)+2)}{\zeta \cdot (\zeta(1-\alpha)+2)} \cdot \sum_{\mat b \in B_{\text{Good, Low}, j}}\hw(\mat b) \\
    = & \frac{\zeta-1}{\zeta} \cdot \sum_{\mat b \in B_{\text{Good, Low}, j}}\hw(\mat b) \\
\end{split}\]
To determine $\hw(\vee_{i \in S}\mathbf{S}_i)$, we can sum the hamming weights for each section in the final vector. Considering only the blocks in $B_{\text{Good, Low}}$, we can write this as:
\[\hw(\vee_{i \in S}\mathbf{S}_i) \geq \sum_j\left(\hw(\bigvee_{\mat b \in B_{\text{Good, Low}, j}}{\mat b})\right)\]
Now composing all of our results and using that $\zeta = \lceil\frac{1}{1-\sqrt{\gamma}}\rceil + 1$:
\[\begin{split}
    \sum_j\left(\hw(\bigvee_{\mat b \in B_{\text{Good, Low}, j}}{\mat b})\right) > & \sum_j\left(\frac{\zeta-1}{\zeta} \cdot \sum_{\mat b \in B_{\text{Good, Low}, j}}\hw(\mat b)\right) \\
    = & \frac{\zeta-1}{\zeta} \cdot \sum_{\mat b \in B_{\text{Good, Low}}}\hw(\mat b) \\
    \geq & \frac{\zeta-1}{\zeta} \cdot \frac{\zeta-1}{\zeta} \cdot (1-o(1)) \cdot k|S| \\
    = & \left(\frac{\lceil\frac{1}{1-\sqrt{\gamma}}\rceil}{\lceil\frac{1}{1-\sqrt{\gamma}}\rceil + 1}\right)^2 \cdot (1-o(1)) \cdot k|S| \\
    > & \left(\frac{\frac{1}{1-\sqrt{\gamma}}-1}{\frac{1}{1-\sqrt{\gamma}}}\right)^2 \cdot k|S| \\
    = & \left(\frac{1 - (1-\sqrt{\gamma})}{1}\right)^2 \cdot k|S| \\
    = & \gamma \cdot k|S|.\\
\end{split}\]
Thus, we achieve the desired lower bound with all but $o(1)$ probability.
\end{proof}

\subsubsection{Proof of Non-Compression}
As stated in Section \ref{sec:ass}, $\mathbf{S}$ will also be a $(\gamma, \Theta(\log n))$-non-compressing matrix with high probability. We begin our proof by stating the Kruskal-Katona theorem~\cite{kruskal63,katona87}. At a high level, this relates the number unique sets of size $\zeta-1$ contained in a collection of sets, each of which has size $\zeta$, for arbitrary values of $\zeta$.

\begin{lemma}[Simplified version of the Kruskal-Katona Theorem \cite{kruskal63,katona87}]\label{thm:kk}
    Consider any set system $\mathcal{T}$ where every $T \in \mathcal{T}$ has cardinality $\zeta$, for any constant integer $\zeta > 1$. Define
    \[\mathcal{Q} = \{C : |C| = (\zeta-1) \land (\exists T \in \mathcal{T} \text{ such that } C \subset T)\}\]
    Then $|\mathcal{Q}| \geq \Omega(|\mathcal{T}|^{(\zeta-1)/\zeta})$.
\end{lemma}

\begin{lemma}\label{thm:noncompressing}
    For all constants $\alpha, \delta \in (0, \frac{1}{2})$ and $\gamma \in (\frac{1}{2}, 1)$ and $\beta > 0$, for all large enough $\lambda$, the following holds. Let $\bar{n} = 2^{(\log\lambda)^{1/(1+\alpha)}}$, $\bar{k} = \bar{n}^\delta$, $\ell=(\log\bar{n})^\alpha$, $q=\bar{n}^{-(\delta\ell)/2}$, $\zeta = \lceil\frac{1}{1-\sqrt{\gamma}}\rceil + 1$, and $\rho=2+\beta\zeta$. Let $G \getsr \mathcal{G}(\bar{n}, \frac{1}{2})$, $H \getsr \RP_q^\ell(G)$, and $\mathbf{S} \gets \Encode(H,\zeta,\beta)$. Then $\mathbf{S} \in \mathbb{F}_2^{m \times n}$ is a $(\gamma, \Theta(\log n))$-non-compressing matrix with probability $1 - o(1)$ over the randomness of $\Gen$.
\end{lemma}
\begin{proof}
As argued in Lemma \ref{thm:expanding}, each row of $\mathbf{S}$ is exactly $k$-sparse, where $k = \Theta(\log n)$ with $1 - o(1)$ probability.\\\\
Consider any set system $\mathcal{K} = \{X_i^{(\zeta)} : i \in P\}$, where $P \subseteq \{1, \ldots m\}$ and each $X_i^{(\zeta)}$ is the set of vertices in $H$ for the $\zeta$-clique of $H$ with label $R_i$. Define
\[\mathcal{L} = \{L : |L| = (\zeta-1) \land (\exists X \in \mathcal{K} \text{ such that } L \subset X)\}.\]
In other words, $\mathcal{L}$ is the set of all $(\zeta-1)$-cliques contained in some $\zeta$-clique of $\mathcal{K}$. Since $\gamma \in (\frac{1}{2}, 1)$ is a constant and $\zeta = \lceil\frac{1}{1-\sqrt{\gamma}}\rceil + 1$, we have that $\zeta > 4$. Therefore by Lemma \ref{thm:kk}, we have that $|\mathcal{L}| \geq \Omega(|\mathcal{K}|^{(\zeta-1)/\zeta})$. Expanding this expression, we get:
\[\begin{split}
    |\mathcal{L}| \geq & \Omega(|\mathcal{K}|^{\lceil\frac{1}{1-\sqrt{\gamma}}\rceil/(\lceil\frac{1}{1-\sqrt{\gamma}}\rceil + 1)}) \\
    > & \Omega(|\mathcal{K}|^{(\lceil\frac{1}{1-\sqrt{\gamma}}\rceil-1)/\lceil\frac{1}{1-\sqrt{\gamma}}\rceil}) \\
    = & \Omega(|\mathcal{K}|^{1 - 1/\lceil\frac{1}{1-\sqrt{\gamma}}\rceil}) \\
    \geq & \Omega(|\mathcal{K}|^{1 - 1/(\frac{1}{1-\sqrt{\gamma}})}) \\
    = & \Omega(|\mathcal{K}|^{\sqrt{\gamma}}) \\
\end{split}\]
Now consider the set of row vectors $\mathcal{S}$ for $\mathbf{S}$. Any subset $\mathcal{T} \subset \mathcal{S}$ corresponds to exactly $|\mathcal{T}|$ many distinct $\zeta$-cliques in $H$, and (by the above) at least $\Omega(|\mathcal{T}|^{\sqrt{\gamma}})$ many $(\zeta-1)$-cliques of $H$. By construction, each of these $(\zeta-1)$-cliques corresponds to at least one distinct ``1'' entry in the bitwise-OR of all vectors $\bigvee_{\mat v \in \mathcal{T}} \mat v$. Therefore we have
\[\begin{split}
    \hw\left( \bigvee_{\mat v \in \mathcal{T}} \mat v \right) \geq & \Omega(|\mathcal{T}|^{\sqrt{\gamma}}). \\
\end{split}\]
Since each row vector is exactly $k$-sparse, we also have that any subset $\mathcal{T} \subset \mathcal{S}$ with $|\mathcal{T}| \leq k$ satisfies
\[\begin{split}
    \hw\left( \bigvee_{\mat v \in \mathcal{T}} \mat v \right) \geq & k \geq |\mathcal{T}|. \\
\end{split}\]
Combining these two observations and noting that $k$ is superconstant implies that
\[\begin{split}
    \hw\left( \bigvee_{\mat v \in \mathcal{T}} \mat v \right) \geq & |\mathcal{T}|^{\gamma}. \\
\end{split}\]
holds for sufficiently large $\lambda$.
\end{proof}

\subsubsection{Semantic Security}
We first show that the scheme is a weak PKE.

\begin{lemma}[Weak Security] \label{lem:weaksecurity}
        Assume that Conjecture \ref{con:plantedclique} holds for some constants $\alpha, \delta \in (0, \frac{1}{2})$, and assume that Conjecture \ref{con:dwdexpanderkosher} holds for some constant $\gamma \in (\frac{1}{2}, 1)$. Then for all large enough $\lambda$, the following is true. Set $\bar{n} = 2^{(\log\lambda)^{1/(1+\alpha)}}$, $\bar{k} = \bar{n}^\delta$, $\ell=(\log\bar{n})^\alpha$, $q=\bar{n}^{-(\delta\ell)/2}$, $\zeta = \lceil\frac{1}{1-\sqrt{\gamma}}\rceil + 1$, $\beta = \zeta \cdot (\zeta(1-\alpha)+2)$, and $\rho=2+\beta\zeta$. For all \text{poly}$(\lambda)$-size algorithms $\cA$,
        \begin{align*}
            & \Big \lvert \Pr \left[ (\pk, \sk) \from \Gen(1^\lambda); \cA(\pk, \Enc(\pk, 0)) = 1,  \right ]\\
            & \qquad - \Pr \left[ (\pk, \sk) \from \Gen(1^\lambda);  \cA(\pk, \Enc(\pk, 1)) = 1 \right ] \Big \rvert \leq \frac{12}{13},
        \end{align*}
        where the probability is taken over the coins of $\Encode$, $\Gen$, and $\Enc$.
\end{lemma}

\begin{proof}
    We will use standard hybrid arguments.
    \begin{mdframed}
        Hybrid $\cH_0(1^\lambda)$:
    \begin{enumerate}
        \item $G \getsr \cG(\bar{n},\frac{1}{2},\bar{k})$.
        \item $H \getsr \RP^\ell_p(G)$.
        \item $\mathbf{S} \gets \Encode(H,\zeta,\beta)$. Let $\mathbf{S}$ have $m$ rows and $n$ columns.
        \item Let $\mathbf{A} \in \mathbb{F}_p^{m \times n}$ be the matrix where we replace each nonzero entry in $\mathbf{S}$ with a random element in $\mathbb{F}_p^*$.
        \item $\pk \coloneqq \mat{A}$.
        \item $\ct \from \Enc(\pk, 0)$ where
        \[\ct \xleftarrow{} \mat{A}\mat{s}+\mat{e}.\]
        Here $\mat{s} \getsr \bF_p^{n}$ and we sample $\mat{e}$ as:
        \[\mathbf{e}_i=\begin{cases} \text{Random element of $\mathbb{F}_{p}^*$, with probability $\eta= (\log n)^{-\varsigma}$. } \\ 0, \text{ otherwise.} \end{cases}\]
        \item Output $(\pk, \ct)$.
    \end{enumerate}
    \end{mdframed}

    \begin{mdframed}
        Hybrid $\cH_{0, \$}(1^\lambda)$:
    \begin{enumerate}
        \item \textcolor{orange}{\underline{$G \getsr \cG(\bar{n},\frac{1}{2})$.}}
        \item $H \getsr \RP^\ell_p(G)$.
        \item $\mathbf{S} \gets \Encode(H,\zeta,\beta)$. Let $\mathbf{S}$ have $m$ rows and $n$ columns.
        \item Let $\mathbf{A} \in \mathbb{F}_p^{m \times n}$ be the matrix where we replace each nonzero entry in $\mathbf{S}$ with a random element in $\mathbb{F}_p^*$.
        \item $\pk \coloneqq \mat{A}$.
        \item $\ct \from \Enc(\pk, 0)$ where
        \[\ct \xleftarrow{} \mat{A}\mat{s}+\mat{e}.\]
        Here $\mat{s} \getsr \bF_p^{n}$ and we sample $\mat{e}$ as:
        \[\mathbf{e}_i=\begin{cases} \text{Random element of $\mathbb{F}_{p}^*$, with probability $\eta= (\log n)^{-\rho}$. } \\ 0, \text{ otherwise.} \end{cases}\]
        \item Output $(\pk, \ct)$.
    \end{enumerate}
    \end{mdframed}

    \begin{mdframed}
        Hybrid $\cH_{1, \$}(1^\lambda)$:
    \begin{enumerate}
        \item $G \getsr \cG(\bar{n},\frac{1}{2})$.
        \item $H \getsr \RP^\ell_p(G)$.
        \item $\mathbf{S} \gets \Encode(H,\zeta,\beta)$. Let $\mathbf{S}$ have $m$ rows and $n$ columns.
        \item Let $\mathbf{A} \in \mathbb{F}_p^{m \times n}$ be the matrix where we replace each nonzero entry in $\mathbf{S}$ with a random element in $\mathbb{F}_p^*$.
        \item $\pk \coloneqq \mat{A}$.
        \item \textcolor{orange}{\underline{$\ct \from \Enc(\pk, 1)$ where}
        \[\underline{\ct \xleftarrow{} \mat{b}}\]
        \underline{where $\mat{b} \getsr \bF_p^{m}$.}}
        \item Output $(\pk, \ct)$.
    \end{enumerate}
    \end{mdframed}

    \begin{mdframed}
        Hybrid $\cH_{1}(1^\lambda)$:
    \begin{enumerate}
        \item \textcolor{orange}{\underline{$G \getsr \cG(\bar{n},\frac{1}{2},\bar{k})$.}}
        \item $H \getsr \RP^\ell_p(G)$.
        \item $\mathbf{S} \gets \Encode(H,\zeta,\beta)$. Let $\mathbf{S}$ have $m$ rows and $n$ columns.
        \item Let $\mathbf{A} \in \mathbb{F}_p^{m \times n}$ be the matrix where we replace each nonzero entry in $\mathbf{S}$ with a random element in $\mathbb{F}_p^*$.
        \item $\pk \coloneqq \mat{A}$.
        \item $\ct \from \Enc(\pk, 1)$ where
        \[\ct \xleftarrow{} \mat{b}\]
        where $\mat{b} \getsr \bF_p^{m}$.
        \item Output $(\pk, \ct)$.
    \end{enumerate}
    \end{mdframed}

    Before analyzing the hybrids, note that $\lambda = \bar{n}^{\Theta(\ell)}$ and $\lambda = n^{\Theta(1)}$ with very high probability.

    \begin{claim}[$\cH_0 \approx_{\frac{1}{3}} \cH_{0, \$}$]
    \label{lemma:h0h0r}
        Assume that Conjecture \ref{con:plantedclique} holds for some constants $\alpha, \delta \in (0, \frac{1}{2})$. For all sufficiently large $\lambda$, and for all constant $\gamma \in (\frac{1}{2}, 1)$, the following is true. Set the parameters as in the lemma statement. For all poly$(\lambda)$-size algorithms $\cA$, 
        \[\Abs{\Pr \left[ \cA (1^\lambda,\cH_0(1^\lambda)) = 1 \right] - \Pr \left[ \cA (1^\lambda, \cH_{0, \$}(1^\lambda)) = 1 \right] } \leq \frac{1}{3}.\]
    \end{claim}
    \begin{proof}
        The only difference between $\cH_0$ and $\cH_{0, \$}$ is the presence or absence of a planted clique. Therefore, a poly$(\lambda)$-size algorithm $\cA$ for differentiating between the two hybrids implies a poly$(\lambda)$-size algorithm $\cB$ for the planted clique problem.\\\\
        Specifically, $\cB$ uses its input graph $G^*$ in place of graph $G$ from Step 1, then follows steps 2 through 7 (which are the same in both hybrids and can be completed in poly$(\lambda)$ time), and finally gives the result as input to $\cA$. The output from $\cA$ is used as the output for $\cB$. Clearly by our choice of parameters, the distinguishing advantage of $\cB$ is identical to that of $\cA$.

    \end{proof}

 \begin{claim}[$\cH_{0,\$} \approx_{\frac{1}{4} + o(1)} \cH_{1, \$}$]
    \label{lemma:h0rh1r}
        Assume that Conjecture \ref{con:dwdexpanderkosher} holds for some constant $\gamma \in (\frac{1}{2}, 1)$. For all sufficiently large $\lambda$, and for all constants $\alpha, \delta \in (0, \frac{1}{2})$, the following is true. Set the parameters as in the lemma statement. For all poly$(\lambda)$-size algorithms $\cA$,
        \[\Abs{\Pr \left[ \cA (1^\lambda,\cH_{0,\$}(1^\lambda)) = 1 \right] - \Pr \left[ \cA (1^\lambda, \cH_{1, \$}(1^\lambda)) = 1 \right] } \leq \frac{1}{4} + o(1).\]
    \end{claim}
    \begin{proof}
    We have already proved that when we start with a graph $G \getsr \cG(\bar{n},\frac{1}{2})$, then the matrix $\mat{A}$ generated as per our encoding procedure above satisfies the preconditions of Conjecture~\ref{con:dwdexpanderkosher}, namely Lemma \ref{thm:expanding}, with $1 - o(1)$ probability. Assuming this event occurs, it is easy to see that any adversary that can distinguish between Hybrids $\cH_{0,\$}$ and $\cH_{1, \$}$ with advantage greater than $\frac{1}{4}$ can be transformed into one that can distinguish between $(\mat A, \mat A \mat s + \mat e)$ and $(\mat A, \$)$ with advantage better than $\frac{1}{4}$. This will contradict Conjecture~\ref{con:dwdexpanderkosher}.

    \end{proof}

      \begin{claim}[$\cH_{1, \$} \approx_{\frac{1}{3}} \cH_{1}$]
    \label{lemma:h1rh1}
        Assume that Conjecture \ref{con:plantedclique} holds for some constants $\alpha, \delta \in (0, \frac{1}{2})$. For all sufficiently large $\lambda$, and for all constant $\gamma \in (\frac{1}{2}, 1)$, the following is true. Set the parameters as in the lemma statement. For all poly$(\lambda)$-size algorithms $\cA$,
        \[\Abs{\Pr \left[ \cA (1^\lambda,\cH_{1,\$}(1^\lambda)) = 1 \right] - \Pr \left[ \cA (1^\lambda, \cH_{1}(1^\lambda)) = 1 \right] } \leq \frac{1}{3}.\]
    \end{claim}
    \begin{proof}
        This follows identically from the proof of Claim~\ref{lemma:h0h0r}.
    \end{proof}
    Combining the above three claims, and accounting for the fact that $\lambda = \bar{n}^{\Theta(\ell)}$ and $\lambda = n^{\Theta(1)}$ with very high probability, we get
        \begin{align*}
            & \Big \lvert \Pr \left[ (\pk, \sk) \from \Gen(1^\lambda); \cA(\pk, \Enc(\pk, 0)) = 1,  \right ]\\
            & \qquad - \Pr \left[ (\pk, \sk) \from \Gen(1^\lambda);  \cA(\pk, \Enc(\pk, 1)) = 1 \right ] \Big \rvert \leq \frac{11}{12} + o(1) < \frac{12}{13}.
        \end{align*} 
\end{proof}

We are now ready to prove our main theorem.

\begin{theorem}
\label{thm:scheme}
    Assuming Conjecture \ref{con:plantedclique} and Conjecture \ref{con:dwdexpanderkosher} both hold, there exists a semantically secure public key encryption scheme.
\end{theorem}

\begin{proof}
    Clearly $o(1) < \frac{1-\sqrt{12/13}}{2} $. Thus, by Lemma \ref{lem:weaksecurity}, the encryption scheme constructed above is an $\left(o(1),\frac{12}{13}\right)$ weak PKE. The theorem then follows from Lemma~\ref{lem:amp}.
\end{proof}

%% file: Properties.tex
\section{Useful Properties of Random Graphs} \label{sec:randomgraphs}

In this section, we will prove certain properties of random graphs that will eventually be helpful towards proving Lemma~\ref{lem:highdegree}.

Throughout, we will consider $G \coloneqq (V_G, E_G) \getsr \mathcal{G}(n, \frac{1}{2})$ and $H \coloneqq (V_H, E_H) \getsr \RP_q^{\ell}(G)$. Let $\bar{n} := |V_G|$, $q := \bar{n}^{-(\delta\ell)/2}$, and $\ell := (\log \bar{n})^\alpha$, for any positive constants $\alpha, \delta \in (0, \frac{1}{2})$. Recall that we can interpret every node $v \in V_H$ as corresponding to a set $U_v$ of $\ell$ many nodes in $G$, and interpret any edge $e \in E_H$ connecting nodes $(v,w) \in V_H$ as corresponding to a set of edges $W_{e}=K(U_v \cup U_w)$. Here $K(X)$ denotes the clique induces by vertices in the set $X$.

\begin{definition}[Vertex and Edge Mappings] \label{defn:mappings}
 For a subset of vertices $S \subseteq V_H$, define \[G^V(S) \coloneqq \bigcup_{v \in S} U_v,\] where $U_i$ is as defined above, and for a subset of edges $T \subseteq E_H$, define \[G^E(T) \coloneqq \bigcup_{e \in T} W_e,\] where $W_e$ is as defined above.

    In other words $G^V(S)$ is the union of vertices in $G$ for each $\ell$-subset corresponding to a vertex of $S$ and $G^E(T)$ is the union of edges in $G$ for each clique in $G$ corresponding to an edge of $T$.
\end{definition}

\begin{lemma} \label{lem:vertexdensity}
     For all constants $\alpha, \delta \in (0, \frac{1}{2})$, there exists a constant $c = c(\delta)$ such that the following is true. Let $\bar{n}$ be any parameter. Set $\ell=(\log\bar{n})^\alpha$ and $q=\bar{n}^{-(\delta\ell)/2}$. Let $G$ be any graph with $\bar{n}$ many vertices, and sample $H \getsr \RP_q^\ell(G)$. With all but $o(1)$ probability over the randomness of $\RP_q^\ell$ we have that, for all $S \subset V_H$ with $\vert S \vert \leq \bar{n}^{o(1)}$, $|G^V(S)| > c\ell|S|$.
\end{lemma}

\begin{proof}
    Suppose for contradiction that there exists some $S \subset V_H$ with $|S| = \bar{n}^{o(1)}$ and such that $|G^V(S)| \leq c\ell|S|$. This implies that there exists an $x =\bar{n}^{o(1)}$ and some subset $S' \subset V_G$ of size $y\leq c\ell x$ 
    with the property that at least $x$ many vertices $v$ in $V_H$ have $U_v \subset S'$. If this event occurs for some $y < c\ell x$, this automatically implies that the event also occurs for all larger values of $y$, so throughout the proof we will assume $y = c\ell x$.\\\\Let $N_x$ denote the expected number of such subsets $S'$ for a fixed choice of $x$, and $N=\sum_{x \leq \bar{n}^{o(1)}} N_x$.\\\\ Observe that there are $\binom{\bar{n}}{y}$ many ways to choose a set $S'$ of $y$ many nodes in $G$. Furthermore, $\binom{y}{\ell}$ is the total number of $\ell$-sized subsets of a given $S'$, so there are $\binom{\binom{y}{\ell}}{x}$ many ways to choose $x$ many $\ell$-sized subsets. The probability that all $x$ of these were actually sampled by $\RP_q^{\ell}$ (in other words, the probability that they all exist as vertices of $H$) is $q^x$. \\\\ Take $c = \delta/5$. We show by linearity of expectation that
\begin{align*}
    \bE[N_x] \leq & 
        \binom{\bar{n}}{y}\cdot \binom{\binom{y}{\ell}}{x} \cdot q^x \\
         < & \bar{n}^{y} \cdot y^{x\ell} \cdot p^q \\
         \leq & 2^{y\log(\bar{n}) + x\ell\log(y) - \frac{1}{2}x\ell\delta\log(\bar{n})} \\
        \leq & 2^{y\log(\bar{n}) + x\ell \cdot (\log(y) - \frac{1}{2}\delta\log(\bar{n}))} \\
        \leq & 2^{y\log(\bar{n}) + x\ell \cdot (\log x+\log \ell+\log c - \frac{1}{2}\delta\log(\bar{n}))} \tag{$y = cx\ell$}\\
        \leq & 2^{y\log(\bar{n}) + x\ell \cdot (o(1) - \frac{1}{2}\delta)\log(\bar{n})} \tag{$\ell\leq\bar{n}^{o(1)}, x\leq\bar{n}^{o(1)}$} \\
         \leq & 2^{- x\ell \cdot \frac{\delta}{4}\log(\bar{n})}=2^{-x\cdot\omega(\log \bar{n})} \tag{$y=cx\ell$, and $c<\delta/4$}
\end{align*} Taking a union bound over all possible values of $x$ gives \[\bE[N]=\sum_{x \leq n^{o(1)}} \bE[N_x] = \sum_{x \leq n^{o(1)}} 2^{-x \omega(\log \bar{n})} \leq {2^{-\omega (\log \bar{n})}}.\] Finally using Markov's inequality, \[\Pr[N \geq 1] \leq \bE[N],\] and therefore $N=0$ with all but $o(1)$ probability. 
    

\end{proof}

\begin{lemma} \label{lem:edgedensity}
    For all constants $\alpha, \delta \in (0, \frac{1}{2})$, there exists a constant $c^* = c^*(\delta)$ such that the following is true. Let $\bar{n}$ be any parameter. Set $\ell=(\log\bar{n})^\alpha$ and $q=\bar{n}^{-(\delta\ell)/2}$. Sample $G \getsr \mathcal{G}(n, \frac{1}{2})$ and $H \getsr \RP_q^\ell(G)$. With all but $o(1)$ probability over the random choice of $G$ and the randomness of $\RP_q^\ell$ we have that, for all subgraphs $S = (V_S, E_S)$ of $H$ with $|V_S| = 2^{o((\log\bar{n})^{2\alpha})}$, $|E_S| < c^*(|V_S|(\log \bar{n})^{1-\alpha})$.
\end{lemma}
\begin{proof}
Assume that the event in Lemma \ref{lem:vertexdensity} occurs. Now suppose for the sake of contradiction that there exists a subgraph $S$ of $H$ with $|V_S| = 2^{o((\log\bar{n})^{2\alpha})}$ and $|E_S| \geq c^* |V_S|(\log \bar{n})^{1-\alpha}$. \\\\ We can actually argue the contradiction for just $|E_S| = c^* |V_S|(\log \bar{n})^{1-\alpha}$. This is because, if there exists a subgraph $S$ with $|E_S| > c^* |V_S|(\log \bar{n})^{1-\alpha}$, we can arbitrarily delete edges of $S$ to get a subgraph $S'$ with $|E_{S'}| = c^* |V_{S'}|(\log \bar{n})^{1-\alpha}$ and $|V_{S'}| = |V_S|$.\\\\
Let $x \coloneqq |V_S|$ and $yx \coloneqq |E_S|$ (which implies that $y = c^*(\log\bar{n})^{1-\alpha}$). Let $N_S(v_S)$ denote the open neighborhood of $v_S \in V_S$ within the subgraph $S$. 
Without loss of generality, we may assume that the minimum degree of each vertex in $S$ is $\lfloor y\rfloor = \lfloor c^*(\log \bar{n})^{1-\alpha}\rfloor$. If not, since $y$ is independent of $|S|$, we can repeatedly delete each vertex of degree less than $\lfloor y\rfloor$ to get a smaller subgraph $S'$ with minimum degree $\lfloor y\rfloor$ and $x' < x$ many vertices. For ease of exposition, we will actually just assume a minimum degree of $\frac{y}{2}$, since $\frac{y}{2} \leq \lfloor y\rfloor$ for any $y \geq 1$. \\\\
By Lemma \ref{lem:vertexdensity}, there exists some constant $c = f(\delta)$ such that \[|G^V(N_S(v_S))| > c\ell|N_S(v_S)| \geq c\ell \frac{y}{2}\] for all $v_S \in V_S$. \\\\
Define a mapping $\phi: G^V(V_S) \rightarrow V_S$ such that, for all $v_G \in G^V(V_S)$, we have that $\phi(v_G) = v_S$ for some $v_S \in V_S$ such that $v_G \in U_{v_S}$. There may be multiple choices of $v_S$ for each $v_G$; just choose from these arbitrarily. By the definition of $G^V(\cdot)$, such a mapping always exists. \\\\
Let $E_\phi(v_G) \subset G^E(E_S)$ be the set of edges between a vertex $v_G \in G^V(V_S)$ and the vertices in $G^V(N_S(\phi(v_G)))$. Note that $v_G$ must have an edge to every vertex in $G^V(N_S(\phi(v_G))) \backslash v_G$ by the definition of the randomized graph product. So $(|G^V(N_S(\phi(v_G)))| - 1)$ is a lower bound on the degree of $v_G \in G^V(V_S)$ with respect to the edges in $G^E(E_S)$. \\\\
We already showed that $|G^V(N_S(\phi(v_G)))| > c\ell \frac{y}{2}$ for all possible $\phi(v_G)$, and therefore this statement holds for all $v_G$. Thus $c\ell\frac{y}{2}-1$ is a strict lower bound for the degree of each $v_G$ with respect to $G^E(E_S)$. (In other words the min degree is at least $c\ell\frac{y}{2}$.) \\\\ 
Applying Lemma \ref{lem:vertexdensity} again, and using that having constant $\alpha \in (0, \frac{1}{2})$ implies that $|V_S| = 2^{o((\log\bar{n})^{2\alpha})} = \bar{n}^{o(1)}$, we know that $|G^V(V_S)| > c\ell|V_S|$.\\\\
Let $z \coloneqq |G^E(E_S)|$. Putting all of this together, and observing that the total number of edges in any graph $Z = (V_Z, E_Z)$ is at least $\frac{1}{2}\cdot|V_Z|\cdot\text{MinDegree}(Z)$, we have \[|G^E(E_S)| \geq \frac{1}{2}\cdot|G^V(V_S)| \cdot c\ell\frac{y}{2} > \frac{1}{4} \cdot c^2 \ell^2 \cdot y|V_S| = \frac{1}{4} \cdot c^2 \ell^2 \cdot yx.\]
We now apply linearity of expectation to show that the number of subgraphs $S$ of $H$ with $|V_S| = 2^{o((\log\bar{n})^{2\alpha})}$ and $|E_S| = c^* |V_S|(\log \bar{n})^{1-\alpha}$ is vanishing by analyzing the edge probabilities in $G$. Let $N_{x}$ be the expected number of subgraphs in $H$ with $x$ nodes and $yx$ edges. Setting $c^* = \frac{15}{c^2}$ and using that the edge probability in $G$ is $\frac12$, we have 
 
    \begin{align*}
        \mathbb{E}[N_{x}] = & \binom{|V_H|}{x} \cdot \binom{\binom{x}{2}}{xy} \cdot 2^{-z}\\
        < & \bar{n}^{x\ell} \cdot x^{2xy} \cdot 2^{-z} \tag{$|V_H| < \bar{n}^\ell$}\\
        < & 2^{x\ell\log\bar{n} + 2xy\log x - \frac{1}{4} c^2\ell^2 yx} \tag{$|E^G(E_S)| > \frac{1}{4} c^2\ell^2 yx$} \\
        = & 2^{x(\ell\log\bar n + 2y\log x - \frac{1}{4} c^2\ell^2 y)} \\
        < & 2^{x(\ell\log\bar n - \frac{1}{5} c^2\ell^2 y)} \tag{$\log x = o(\ell^2)$} \\
        = & 2^{x(\ell\log\bar n - \frac{1}{5} c^2\ell  c^* \log\bar{n})} \tag{$\ell = (\log(\bar{n}))^\alpha$ and $y = c^* ((\log \bar{n})^{1-\alpha})$}\\
        = & 2^{x(\ell\log\bar n - 3 \ell \log\bar{n})} \\
        \leq & 2^{- 2x\ell\log\bar{n}} \\
    \end{align*}
   Finally this gives us 
   \begin{align*}
       \bE[N]=&\sum_{x \leq 2^{o((\log\bar{n})^{2\alpha})}} \bE[N_
   {x}]\\
   <&\sum_{x \leq 2^{o((\log\bar{n})^{2\alpha})}} 2^{- 2x\ell\log\bar{n}}\\
   <& o(1)\\ 
   \end{align*}
Using Markov's inequality, \[\Pr[N \geq 1] \leq \bE[N],\] and therefore $N=0$ with all but $o(1)$ probability (assuming the event in Lemma \ref{lem:vertexdensity} occurred, which also happens with all but $o(1)$ probability).

\end{proof}

\begin{lemma} \label{lem:cliqueratio}
For all constants $\alpha, \delta \in (0, \frac{1}{2})$, there exists a constant $\tilde{c} = \tilde{c}(\delta)$ such that the following is true. Let $\bar{n}$ be any parameter and $\zeta > 1$ be any (not necessarily constant) integer. Set $\ell=(\log\bar{n})^\alpha$ and $q=\bar{n}^{-(\delta\ell)/2}$. Sample $G \getsr \mathcal{G}(\bar{n}, \frac{1}{2})$, and sample $H \getsr \RP_q^\ell(G)$. With all but $o(1)$ probability over the random choice of $G$ and the randomness of $\RP_q^\ell$ we have that, for all sets $S$ of $2^{o((\log\bar{n})^{2\alpha})}$ many $\zeta$-cliques in $H$, there exists a set $S'$ of $(\zeta-1)$-cliques such that $\vert S' \vert \geq \frac{\tilde{c} \zeta}{(\log\bar{n})^{1-\alpha}} |S|$ and \[\forall C \in S', \exists K \in S \text{ such that } C \subset K\] 
 

\end{lemma}

\begin{proof}
    We give a proof by induction. Throughout the proof, we will take $S'$ to be the set of all distinct $(\zeta-1)$-cliques contained in some $\zeta$-clique of $S$. \\\\
    For the base case of $\zeta = 2$, the lemma corresponds to lower bounding the ratio of vertices to edges, and so the claim holds by Lemma \ref{lem:edgedensity}. 
    For the inductive case, we will count the number of $(\zeta-1)$-cliques and $\zeta$-cliques that each vertex participates in, and then use this to bound the overall ratio between $\vert S'\vert$ and $\vert S \vert$. All of the inductive cases are deterministic, so they follow immediately if the event in Lemma \ref{lem:edgedensity} occurs.\\\\
    Consider some set $S$ satisfying the condition in the lemma, and let $V_S$ be the union of vertices for all cliques in $S$. We now examine an arbitrary vertex $v \in V_S$. Each distinct $\zeta$-clique $K \in S$ with $v \in K$ contains a unique $(\zeta-1)$-clique, namely $C \coloneqq K\setminus \{v\}$. Denote the set of all such $(\zeta-1)$-cliques to be $S'_v$. Note that $|S'_v| \leq |S|$, since each $K \in S$ contributes at most one $(\zeta-1)$-clique to $S'_v$. This implies that that the inductive hypothesis also bounds the ratio of $(\zeta-1)$-cliques to $(\zeta-2)$-cliques in $S'_v$.\\\\Observe that $v$ participates in exactly $|S'_v|$ many $\zeta$-cliques. But also, $v$ participates in exactly as many $(\zeta-1)$-cliques as there are $(\zeta-2)$-cliques fully contained in some $(\zeta-1)$-clique $C \in S'_v$. Denote this set of $(\zeta-2)$-cliques to be $S''_v$. Using the inductive hypothesis, $|S''_v| \geq \tilde{c} \cdot (\zeta-1)\cdot |S'_v|/(\log\bar{n})^{1-\alpha}$.\\\\Now consider the sum over all counts for $\zeta$-cliques. We have that \[\sum_{v \in V_S}|S'_v| = \zeta \cdot |S|\] since each $\zeta$-clique is counted exactly $\zeta$ many times. By the same argument, \[\sum_{v \in V_S}|S''_v| = (\zeta-1) \cdot |S'|.\]
    Composing these results yields the lemma:
    \[\begin{split}
        \vert S'\vert = & \frac{1}{\zeta-1}\sum_{v \in V_S}|S''_v| \\
        \geq & \frac{1}{\zeta-1}\sum_{v \in V_S}(\tilde{c} \cdot (\zeta-1) \cdot |S'_v|/(\log\bar{n})^{1-\alpha}) \\
        = & \frac{\tilde{c}}{(\log\bar{n})^{1-\alpha}}\sum_{v \in V_S}|S'_v| \\
        = & \frac{\tilde{c}\zeta}{(\log\bar{n})^{1-\alpha}}\vert S\vert
    \end{split}\]
\end{proof}

\begin{corollary} \label{corr:zetacount}
    Let $\bar{n} \in \bN$ be any parameter, $\zeta > 1$ be any (not necessarily constant) integer, and $\alpha, \delta \in (0, \frac{1}{2})$ be any constants. Set $\ell=(\log\bar{n})^\alpha$ and $q=\bar{n}^{-(\delta\ell)/2}$. Sample $G \getsr \mathcal{G}(\bar{n}, \frac{1}{2})$, and sample $H \getsr \RP_q^\ell(G)$. With all but $o(1)$ probability over the random choice of $G$ and the randomness of $\RP_q^\ell$, the following occurs. Any subgraph $H' = (V_{H'}, E_{H'})$ of $H$ with $|V_{H'}| = 2^{o((\log\bar{n})^{2\alpha})}$ contains at most $O(\frac{|V_{H'}|}{\zeta}(\log\bar{n})^{(\zeta-1)(1-\alpha)})$ many $\zeta$-cliques.
\end{corollary}

\begin{proof}
    The proof follows by iteratively applying Lemma \ref{lem:cliqueratio}.
\end{proof}

\begin{definition}[$(\Delta, S)$-High Degree]\label{def:highdegree} A $(\zeta-1)$-clique is said to be \emph{$(\Delta, S)$-high degree} if it is contained in at least $\Delta$ many distinct $\zeta$-cliques of $S$, where $S$ is a set of $\zeta$-cliques.
\end{definition}

\begin{lemma} \label{lem:highdegree}
    Let $\bar{n} \in \mathbb{N}$ be any parameter, $\zeta > 1$ be any (not necessarily constant) integer, and $\alpha, \delta \in (0, \frac{1}{2})$ be any constants. Set $\ell=(\log\bar{n})^\alpha$, $q=\bar{n}^{-(\delta\ell)/2}$, and $\Delta = (\log\bar{n})^{\zeta(1-\alpha)}$. Sample $G \getsr \mathcal{G}(\bar{n}, \frac{1}{2})$, and sample $H \getsr \RP_q^\ell(G)$. With all but $o(1)$ probability over the random choice of $G$ and randomness of $\RP_q^\ell$, the following occurs. For any set $S$ of $2^{o((\log\bar{n})^{2\alpha})}$ many $\zeta$-cliques of $H$, define $T \subset S$ as \[T=\{K \in S: K \text{ has as a subgraph more than one $(\Delta, S)$-high degree $(\zeta-1)$-clique}\}.\] Then $\vert T \vert=o(1) \cdot \vert S \vert$.

    
\end{lemma}

\begin{proof}
    Consider any set $S$ satisfying the condition in the lemma. By the pigeonhole principle, and the fact that each $\zeta$-clique contains exactly $\zeta$ many $(\zeta-1)$-cliques, at most $\zeta|S|/\Delta$ many $(\zeta-1)$-cliques will be $(\Delta, S)$-high degree. Denote this set of $(\zeta-1)$-cliques to be $S'_{\text{High}}$. We now bound the number of $\zeta$-cliques $K \in S$ that contain at least two $(\zeta-1)$-cliques in $S'_{\text{High}}$.\\\\
    Let $V_{\text{High}} \subset V_H$ be the set of all vertices $v$ such that there exists $C \in S'_{\text{High}}$ with $v \in C$. Applying a naive upper bound, we have that $\vert V_{\text{High}}\vert \leq (\zeta-1)|S'_{\text{High}}|$, which is realized when all of the $(\zeta-1)$-cliques in $S'_{\text{High}}$ are vertex-disjoint. Applying Corollary \ref{corr:zetacount}, we see that at most \[O(\frac{\vert V_{\text{High}}\vert}{\zeta} \cdot (\log\bar{n})^{(\zeta-1)(1-\alpha)}) = O((|S|/\Delta) \cdot (\log\bar{n})^{(\zeta-1)(1-\alpha)}) = O(|S|/(\log\bar{n})^{1-\alpha})\] many $\zeta$-cliques can possibly span $V_{\text{High}}$ (even ignoring whether they are in $S$). The remaining $\zeta$-cliques in $S$ must have at least one vertex outside of $V_{\text{High}}$. But if at least one vertex of a $\zeta$-clique $K \in S$ is outside, then all but one of the $(\zeta-1)$-cliques $C \subset K$ will also have a vertex outside of $V_{\text{High}}$. Since $V_{\text{High}}$ completely covers all high degree $(\zeta-1)$-cliques, this implies that only one $(\zeta-1)$-cliques in $K$ could be high degree.\\\\
    Therefore the total number of all $\zeta$-cliques in $S$ that have more than one $(\Delta, S)$-high degree $(\zeta-1)$-clique is $O(\vert S \vert/(\log\bar{n})^{1-\alpha}) = o(1)\cdot \vert S\vert$.
\end{proof}

%% file: Evidence.tex
\section{Evidence for the Noisy $k\mbox{-}\mathsf{LIN}$ Conjecture}
\label{sec:evidence}


Here we discuss some lower bounds about the special case of $k$-$\mathsf{XOR}$, i.e., $k$-$\mathsf{LIN}$ over $\bF_2$. We refer readers to a recent work by Kocurek and Manohar~\cite{kocurekstrong} for lower bounds for $k$-$\mathsf{Lin}$ over larger fields.

\paragraph{Computational Indistinguishability \textnormal{(Adapted from~\cite{applebaum10})}.} 

We say that a distribution \( \cD \) over \( \{0,1\}^{m} \) \( \varepsilon \)-fools a class \( \mathcal{F} \) of boolean functions over \( \{0,1\}^{m} \) if for every \( f \in \mathcal{F} \) we have:
\[
\left| \Pr[f(\cD) = 1] - \Pr[f(U_{m}) = 1] \right| \leq \varepsilon.
\] Here $U_{m}$ is the uniform distribution over $\bF_2^{m}$. 
Let $\cD_{\mat{A},(\log n)^{-\rho}}$ be the distribution induced by a noisy $k\mbox{-}\mathsf{XOR}$ sample, where $\mat A \in \mathbb{F}_2^{m \times n}$. We import the following theorem about the unconditional hardness of the noisy $k\mbox{-}\mathsf{XOR}$ conjecture from the work of Applebaum, Barak and Wigderson.

\begin{lemma}[Theorem 9.1 in the full version of~\cite{applebaum10}]
Let \( \mat{A} \in \mathbb{F}_2^{m \times n} \) be any $(0.51, k, t)$-expanding matrix. 
Then \( \cD_{\mat{A},(\log n)^{-\rho}} \) satisfies the following:

\begin{enumerate}
    \item \( 0 \)-fools \( t \)-wise tests. \quad (\( \cD_{\mat{A},(\log n)^{-\rho}} \) is \( t \)-wise independent.)
    
    \item \( \delta = \frac{1}{2} \cdot \left( 1 - 2(\log n)^{-\rho} \right)^t \)-fools linear tests. \quad (\( \cD_{\mat{A},(\log n)^{-\rho}} \) is \( \delta \)-biased.)
    
    \item \( 8 \cdot \left( 1 - 2(\log n)^{-\rho} \right)^{t/2^{d-1}} \)-fools degree \( d \) polynomials over \( \mathbb{F}_2 \).
\end{enumerate}
\end{lemma}
In this paper, we use the noisy $k\mbox{-}\mathsf{XOR}$ conjecture with $\rho = O(1)$, $k=\Omega(\log n)$, and $t=2^{\log^\alpha n}$. Moreover, we achieve an expansion of, say, $0.99$, which is significantly higher than the $0.51$ threshold in the above theorem. Plugging these in, we can state the following corollary:

\begin{corollary}
Let \( \mat{A} \in \mathbb{F}_2^{m \times n} \) be a \( d \)-sparse matrix that is a \( (0.51, \log n, 2^{\log^\alpha n}) \)-expander for some constant $\alpha<\frac{1}{2}$. Then \( \cD_{\mat{A},(\log n)^{-\rho}} \) satisfies the following:

\begin{enumerate}
    \item \( 0 \)-fools any \( 2^{\log^\alpha n} \)-wise test.
    
    \item \( 2^{-\Omega(2^{\log^\alpha n})} \)-fools linear tests.
    
    \item \( 2^{-\Omega(2^{\log^\alpha n})} \)-fools degree \( d \) polynomials over \( \mathbb{F}_2 \), for any $d \leq 0.99\log^\alpha n $.
\end{enumerate}
\end{corollary}



\paragraph{Sum of Squares Lower Bounds.} Grigoriev \cite{grigoriev01} showed that a random instance of $3\mbox{-}\mathsf{XOR}$ over $n$ variables with $O(n)$ clauses cannot be refuted by the degree $o(n)$ sum of squares algorithm.
This result was later rediscovered by Schoenebeck \cite{schoenebeck08}, who also observed that it can be used to get lower bounds for random instances of $k\mbox{-}\mathsf{XOR}$ and $k\mbox{-}\mathsf{SAT}$.
In fact, these lower bounds work for any $k\mbox{-}\mathsf{XOR}$ instance that is sufficiently expanding. Barak and Steurer \cite{barak} describe how to derive Grigoriev's result in this generalized sense, with their main lemma being the following.

\begin{lemma}[Paraphrased from Lemma 9 in~\cite{barak}]
    There exists some constant $\bar{c}$ such that the following is true. Let $\phi$ be any max $3\mbox{-}\mathsf{XOR}$ instance whose clauses constitute a $(0.51, 3, t)$-expanding set system. Then the degree $\bar{c}t$ sum of squares algorithm fails to refute $\phi$.
\end{lemma}

In other words, even if only a $0.51$ fraction of the clauses can be satisfied by any assignment of the variables, the degree $\bar{c}t$ sum of squares algorithm fails to show that we cannot satisfy all of the clauses simultaneously.\\\\
Barak and Steurer's proof proceeds independently of the XOR arity, so it directly translates to $k\mbox{-}\mathsf{XOR}$ for any (possibly superconstant) $k$. We can thus state the following corollary.

\begin{corollary}
    There exists some constant $\bar{c}$ such that the following is true. Let $\phi$ be any max $k\mbox{-}\mathsf{XOR}$ instance whose clauses constitute a $(0.51, k, t)$-expanding set system. Then the degree $\bar{c}t$ sum of squares algorithm fails to refute $\phi$.
\end{corollary}

While the above corollary only concerns refutation, not the problem of distinguishing between $(\mat A, \mat A \mat s + \mat e)$ and $(\mat A, \$)$, it is natural to believe that it gives strong evidence for the security of the noisy $k\mbox{-}\mathsf{XOR}$ conjecture.





\paragraph{$\textnormal{AC}^0$ Circuit Lower Bounds.}
We can also apply a result of Braverman regarding $\textnormal{AC}^0$ circuits \cite{braverman08}. We import the following theorem.
\begin{lemma}[From \cite{braverman08}]\label{thm:circuit}
    There exists some function $r(m, \varepsilon, d) = (\log(m/\varepsilon))^{O(d^2)}$ such that the following holds. For all $d = O(1)$ and $m = \Omega(1)$, for all $\textnormal{AC}^0$ circuits $X$ of depth $d$ and size $m$, and for all $r(m, d, \varepsilon)$-wise independent distributions $\cD$ over Boolean vectors, we have that $X$ is $\varepsilon$-fooled by $\cD$.
\end{lemma}

Rearranging values, we get the following corollary.

\begin{corollary}
    Any circuit of depth $d = O(1)$ and size $n^{O(1)}$ is $2^{-2^{\Omega(\log^\alpha n)}}$-fooled by $\cD_{\mat{A},(\log n)^{-\rho}}$.
\end{corollary}

%% file: Search.tex
\section{Using the Search Version of Noisy $k\mbox{-}\mathsf{LIN}$}
\label{sec:search}




\subsection{PKE Scheme}
This construction is very similar to the previous one. The only major change is in the decryption procedure.

\paragraph{Notations.} Recall that we can interpret every node $v \in V_H$ as corresponding to a set $U_v$ of $\ell$ many nodes in $G$, and interpret any edge $e \in E_H$ connecting nodes $(v,w) \in V_H$ as corresponding to a set of edges $W_{e}=K(U_v \cup U_w)$. Here $K(X)$ denotes the clique induces by vertices in the set $X$.

As before, for a subset of vertices $S \subseteq V_H$, define \[G^V(S) \coloneqq \bigcup_{v \in S} U_v,\] where $U_i$ is as defined above, and for a subset of edges $T \subseteq E_H$, define \[G^E(T) \coloneqq \bigcup_{e \in T} W_e,\] where $W_e$ is as defined above.

In other words $G^V(S)$ is the union of vertices in $G$ for each $\ell$-subset corresponding to a vertex of $S$ and $G^E(T)$ is the union of edges in $G$ for each clique in $G$ corresponding to an edge of $T$.

\paragraph{Suggested Parameters.} We use the same parameters as for the first PKE scheme. Let $\alpha, \delta \in (0, \frac{1}{2})$ be constants such that Conjecture \ref{con:plantedclique} holds, and let $\gamma \in (\frac{1}{2}, 1)$ be a constant such that Conjecture \ref{con:sexpander} holds. (As noted before, if Conjecture \ref{con:plantedclique} holds for some $\alpha \geq \frac{1}{2}$, this automatically implies that the conjecture holds for all $\alpha \in (0, \frac{1}{2})$.) Assuming the security parameter is $\lambda$,

\begin{itemize}
    \item Size of initial Erdos-Renyi random graph: $\bar{n} = 2^{(\log\lambda)^{1/(1+\alpha)}}$.
    \item Planted clique size: $\bar{k} = \bar{n}^\delta$.
    \item Graph product parameter for $\RP$: $\ell=(\log\bar{n})^\alpha$.
    \item Node deletion probability for $\RP$: $q=\bar{n}^{-(\delta\ell)/2}$.
    \item Clique size (constant integer) for constructing $k\mbox{-}\mathsf{LIN}$ matrix: $\zeta = \lceil\frac{1}{1-\sqrt{\gamma}}\rceil + 1$.
    \item Exponent (constant) for block size in $k\mbox{-}\mathsf{LIN}$ matrix: $\beta = \zeta \cdot (\zeta(1-\alpha)+2)$.
    \item Exponent (constant) for Bernoulli noise for $k\mbox{-}\mathsf{LIN}$ samples: $\rho=2+\beta\zeta$.
\end{itemize}



\paragraph{Key Generation and Encoding Scheme.}
The algorithms $\Gen(1^{\lambda})$ and $\Encode(H,\zeta,\beta)$ are identical to those for the first PKE scheme.

\paragraph{Encryption.}
\begin{mdframed}
\begin{itemize}
    \item[$\star$] $\ct \from \Enc(1^\secParam, \pk, b \in \{0, 1\})$:

    \begin{enumerate}
        \item Parse $\pk$ as $\mat{A}$. Let $\mat{A}$ have $m$ rows and $n$ columns. 
        \item Sample $\mat{e}$ as:
        \[\mathbf{e}_i=\begin{cases} \text{Random element of $\mathbb{F}_{p}^*$, with probability $\eta= (\log n)^{-\rho}$. } \\ 0, \text{ otherwise.} \end{cases}\]
        \item Sample $\mat{s} \getsr \bF_p^{n}$, and then sample $\mat{y} \getsr \bF_2^{n \lceil\log p\rceil}$ conditioned on $\mathbf{\tilde{s}} \cdot \mat{y} = b$, where $\mathbf{\tilde{s}}$ is any canonical bit representation of $\mathbf{s}$.
        \item Output $\ct \gets (\mat y, \mat b = \mat A \mat s + \mat e)$.
    \end{enumerate}
\end{itemize}
\end{mdframed}

\paragraph{Decryption.} Decryption in this case is significantly more involved than it was for the scheme based on decisional $k\mbox{-}\mathsf{LIN}$. This is because to enable decryption, one needs to recover the \emph{entire} secret $\mat{s}$ with $1 - o(1)$ probability. 
This is a strictly stronger requirement than before.

\begin{mdframed}
\begin{itemize}
    \item[$\star$] $b' \from \Dec(1^\secParam, \pk, \sk, \ct)$:
    \begin{enumerate}
        \item Parse $\pk$ as $\mat{A}$. Let $\mat{A}$ have $m$ rows and $n$ columns.
        \item Parse $\sk$ as $(H, K_H)$.
        \item Parse $\ct$ as $(\mat y, \mat b)$.
        \item Let $\eta$ be any (superconstant) integer satisfying $(\log n)^{\rho-1} \leq \binom{\eta}{\zeta} \leq 2(\log n)^{\rho-1}$.
        \item Recompute all canonical labels $C_j$.
        \item We will solve for a vector $\mat s' \in \bF_p^n$ as follows: Interpret the $\mat s'$ as the concatenation of $c_{H, \zeta -1}$ many blocks, each of size exactly $\log^\beta m$. In other words, $(\mat s')^T =(\mat{b}_{1}\|\cdots\|\mat{b}_{n/\log^\beta m})$, where $(\mat s')^T$ is the transpose of $\mat s'$. For each $j \in [n/\log^\beta m]$, we solve for $\mat b_j$ in the following manner.
        \begin{enumerate}
            \item Let $X_j^{(\zeta-1)}$ be the set of vertices in $H$ for the $(\zeta-1)$-clique labeled as $C_j$. 
            \item Let $K^{(j)} = (\cap_{v \in X_j^{(\zeta-1)}}N_H(v)) \cap {K_H}$.
            \item Let $z \coloneqq \lfloor\frac{\vert{K^{(j)}}\vert}{{\eta - (\zeta-1)}}\rfloor$. Randomly partition $K^{(j)}$ into subsets $K^{(j)}_1, \ldots K^{(j)}_z$ of size exactly $\eta - (\zeta - 1)$ and at most one subset $K^{(j)}_{\text{Leftover}}$ of size less than $\eta - (\zeta - 1)$.
            \item For all $a \in [z]$, let $T_a = K^{(j)}_a \cup X_j^{(\zeta-1)}$.
            \item For all $a \in [z]$, let $\mat x_a \from \mathsf{Solve}(H, T_a, \mat A, \mat b)$.
            \item Set $\mat b_j$ to be the majority vote of the corresponding blocks in the $\mat x_a$'s. If there is no majority vector for $\mat b_j$, set $\mat b_j$ to an arbitrary vector.
        \end{enumerate}
        \item Output $b' \gets \mat y \cdot \mat{\tilde{s'}}$, where $\mat{\tilde{s'}}$ is expressed in the same canonical bit representation as in the algorithm $\Enc$.
    \end{enumerate}
\end{itemize}
\end{mdframed}

Below is the algorithm to solve each system of linear equations.

\begin{mdframed}
\begin{itemize}
    \item[$\star$] $\mat x \from \mathsf{Solve}(H, T, \mat A, \mat b)$:
    \begin{enumerate}
        \item Let $\mat{A}$ have $m$ rows and $n$ columns.
        \item Recompute all canonical labels $R_i$ and $C_j$. Recompute all indicator variables $I_{i, j}$.
        \item For all $\mat{A}_i$, define $X^{(\zeta)}_i$ as the subset of $V_H$ corresponding to the $\zeta$ clique labelled as $R_i$. Define a set $S$ of row labels of $\mat{A}$ as \[S=\left\{i: X^{(\zeta)}_i \subset T\right\}\] 
        \item Split every row of $\mat{A}$ into blocks of length $\log^\beta m$, i.e., $\mat{A}_{i}=(\mat{b}_{i,1}\|\cdots\|\mat{b}_{i,c_{H,\zeta-1}})$ where every $\mat{b}_{i,j}$ has width $\log^\beta m$. Define a set of block labels $S' \subset [c_{H,\zeta-1}]$ such that \[S'=\{j: \exists i \in S \text{ such that } I_{i,j} = 1\}.\]
        \item Let $\tilde{\mat{A}}$ be the submatrix of $\mat{A}$ with rows labeled by $S$ and columns made by concatenating the blocks labeled by $S'$.
        \item Solve the following system of linear equations using Gaussian elimination:\[\tilde{\mat{A}}\mat{x}=\mat b\vert_S,\] where $\mat b\vert_S$ is the vector $\mat b$ restricted to indices in $S$.
        \item If the above step returns a valid solution, then output $\mat x$. Otherwise output $\mat 0$. 
    \end{enumerate}
\end{itemize}
\end{mdframed}

\subsection{Correctness}

We first show that for all $j$, $K^{(j)}$ is very large. In particular, this applies even if the $j$\textsuperscript{th} block of $\mat s'$ corresponds to a $(\zeta-1)$-clique that is \emph{entirely outside} of $K_H$.

\begin{lemma} \label{lem:neighborsize}
    For all constants $\alpha, \delta \in (0, \frac{1}{2})$ and $\beta > 0$, for all constant integers $\zeta > 1$, for all large enough $\lambda$, the following is true. Set $\bar{n} = 2^{(\log\lambda)^{1/(1+\alpha)}}$, $\bar{k} = \bar{n}^\delta$, $\ell=(\log\bar{n})^\alpha$, $q=\bar{n}^{-(\delta\ell)/2}$, and $\rho = 2+\beta\zeta$. With $1 - 2^{-\bar{n}^{\Omega(1)}}$ probability over the random choice of $G$ and the randomness of $\RP_q^\ell$, we have that $|K^{(j)}| > \bar{n}^{\delta\ell/3}$ for all $j$. 
\end{lemma}

\begin{proof}
    Recall that $K^{(j)} = (\cap_{v \in X_j^{(\zeta-1)}}N_H(v)) \cap {K_H}$, where $X_j^{(\zeta-1)}$ is the set of vertices for the $(\zeta-1)$-clique in $H$ labeled $C_j$.
    We will analyze each $K^{(j)}$ in terms of the original graph $G$. \\\\
    Let the set of vertices for the planted clique in $G$ be $K_G$.
    The vertices of $X_j^{(\zeta-1)}$ correspond to a set of at most $\ell \cdot (\zeta-1)$ many vertices in $G$, namely $Y = G^V(X_j^{(\zeta-1)})$. \\\\
    By definition of the randomized graph product, we have that $G^V(K^{(j)}) \subseteq K_G^{(j)}$, where $K_G^{(j)} = (\cap_{u \in Y}N_G(u)) \cap K_G$. Furthermore, during $\RP_q^\ell$, every $\ell$-sized subset of $K_G^{(j)}$ becomes a vertex of $K^{(j)}$ with probability $q$. \\\\
    Our first goal is to give a lower bound on $\vert K_G^{(j)}\vert$ that holds for all $j$ with very high probability. To do this, we find it more convenient to consider a random set of vertices as the set $Y$, and then take a union bound over all possible sets of size up to $\ell \cdot (\zeta-1)$. \\\\
    Fix a random set $Y \subset V_G$, conditioned on $|Y| \leq \ell \cdot (\zeta-1)$, and let $K_G^{Y} = (\cap_{u \in Y}N_G(u)) \cap K_G$. We have that $|K_G| = \bar{k} = \bar{n}^\delta$ by construction, and each vertex of $K_G$ is in the neighborhood of all vertices in $Y$ with probability at least $2^{-|Y|}$, over the random choice of $G$ and the randomness of $Y$. (The true probability is slightly greater since some vertices of $Y$ may happen to fall inside of the planted clique.) Furthermore, this event occurs independently for each vertex of $K_G$. We can thus lower bound the probability that $|K_G^{Y}| \geq \bar{n}^{0.9\delta}$ using a Chernoff bound.
    Since $\mathbb{E}[|K_G^Y|] \geq \bar{n}^{\delta}2^{-|Y|} > 2\bar{n}^{0.9\delta}$ when $\bar{n}$ is sufficiently large and $\alpha \in (0, \frac{1}{2})$, we get
    \begin{align*}
        \Pr[|K_G^{Y}| \geq \bar{n}^{0.9\delta}] & = 1 - \Pr[|K_G^{Y}| < \bar{n}^{0.9\delta}] \\
        & \geq 1 - \Pr[|K_G^{Y}| < \frac{1}{2}\mathbb{E}[|K_G^{Y}|]] \\
        & \geq 1 - 2^{-\bar{n}^{\Omega(1)}} \\
    \end{align*}
    This lower bound also holds with $1 - 2^{-\bar{n}^{\Omega(1)}}$ probability after taking a union bound over all subsets $Y \subset V_G$ such that $|Y| \leq \ell \cdot (\zeta - 1)$, since there are less than $\bar{n}^{\ell \cdot (\zeta - 1)}$ such subsets. So with probability $1 - 2^{-\bar{n}^{\Omega(1)}}$, we have that $|K_G^{(j)}| \geq \bar{n}^{0.9}$ holds for all $j$. \\\\
    Fix a random choice of $j$. We now lower bound the number of $\ell$-sized subsets of $K_G^{(j)}$ that were sampled by $\RP_q^\ell(G)$. The number of such subsets is $|K^{(j)}|$. We apply a Chernoff bound again, and use that $\mathbb{E}[|K^{(j)}|\text{ $\Big\vert$ }|K_G^{(j)}| = \bar{n}^{0.9\delta}] > 2\bar{n}^{\ell\delta/3}$.
    \begin{align*}
        \Pr[|K^{(j)}| \geq \bar{n}^{\ell\delta/3}] & \geq \Pr[|K^{(j)}| \geq \bar{n}^{\ell\delta/3}\text{ $\Big\vert$ }K_G^{(j)} \geq \bar{n}^{0.9\delta}] \cdot (1 - 2^{-\bar{n}^{\Omega(1)}}) \\
        & \geq (1 - \Pr[|K^{(j)}| < \frac{1}{2}\mathbb{E}[|K^{(j)}|\text{ $\Big\vert$ }|K_G^{(j)}| \geq \bar{n}^{0.9\delta}]]) \cdot (1 - 2^{-\bar{n}^{\Omega(1)}}) \\
        & > (1 - \Pr[|K^{(j)}| < \frac{1}{2}\mathbb{E}[|K^{(j)}|\text{ $\Big\vert$ }|K_G^{(j)}| = \bar{n}^{0.9\delta}]]) \cdot (1 - 2^{-\bar{n}^{\Omega(1)}}) \\
        & \geq 1 - 2^{-\bar{n}^{\Omega(1)}}
    \end{align*}

Applying a union bound over all choices of $j$ yields the lemma.
\end{proof}

We now show that decryption succeeds with very high probability.

\begin{lemma}
    For all constants $\alpha, \delta \in (0, \frac{1}{2})$ and $\beta > 0$, for all constant integers $\zeta > 1$, for all large enough $\lambda$, the following is true. Set $\bar{n} = 2^{(\log\lambda)^{1/(1+\alpha)}}$, $\bar{k} = \bar{n}^\delta$, $\ell=(\log\bar{n})^\alpha$, $q=\bar{n}^{-(\delta\ell)/2}$, and $\rho=2+\beta\zeta$. Then for all $b \in \{0, 1\}$, 
    \[\Pr_{(\pk, \sk) \from \Gen(1^{\bar{n}})} [\Dec(\sk, \Enc(\pk, b)) = b) ] \geq 1-o(1),\] where the probability is over the coins of $\Encode$, $\Gen$, $\Enc$, and $\Dec$.
\end{lemma}

\begin{proof}
    Assume that the event in Lemma \ref{lem:neighborsize} occurs. Consider a single block $\mat b_j$ of $\mat s'$. By Lemma \ref{lem:neighborsize}, when solving for $\mat b_j$, we take the majority vote of the solutions to at least $\bar{n}^{\delta\ell/3 -o(1)}$ many systems of linear equations. Denote this set of systems to be $\mathcal{E}$. \\\\
    Each system $E_a \in \mathcal{E}$ is derived from a set $K^{(j)}_a \cup X_j^{(\zeta-1)}$ of vertices in $H$. Specifically, the equations are defined by the $\zeta$-cliques in $K^{(j)}_a \cup X_j^{(\zeta-1)}$. Because every $K^{(j)}_a$ is vertex-disjoint, we have that every $\zeta$-clique in $K^{(j)}_a \cup X_j^{(\zeta-1)}$ is distinct from every $\zeta$-clique in $K^{(j)}_{a'} \cup X_j^{(\zeta-1)}$, for all $a \neq a'$. So for every $E_a, E_{a'} \in \mathcal{E}$ with $a \neq a'$, every equation in $E_a$ is distinct from every equation in $E_{a'}$. (In fact, all of the $(\zeta-1)$-cliques are also distinct, except for the one labeled $C_j$, which is common across all $E_a \in \mathcal{E}$.) \\\\
    Using that $K^{(j)}$ was partitioned randomly, we know that each $E_a \in \mathcal{E}$ will give the correct value for all variables in its system with probability $1-o(1)$, via the same proof technique as in Lemma \ref{lem:correct}. In fact, the probability is only over the randomness of the error vector restricted the that specific set of equations. This means that every $E_a$ gives the correct value for $\mat b_j$ independently with probability $1-o(1)$. By a Chernoff bound, the probability that more than half of the systems give the correct value is more than $1 - 2^{-\bar{n}^{\Omega(1)}}$. If this occurs, then the majority vote must give the correct value of $\mat b_j$ regardless of the values given by the remaining systems.\\\\
    The lemma follows by taking a union bound over all blocks of $\mat s'$, multiplying by the probability that the even in Lemma \ref{lem:neighborsize} occurs, and noting that recovering $\mat s' = \mat s$ implies $b' = b$.
\end{proof}

\subsection{Security}
Security follows from nearly the same hybrid argument as in Section \ref{sec:pke}, but with an additional appeal to the Goldreich-Levin Theorem~\cite{goldreich89}.

\begin{lemma}[Weak Security]
        Assume that Conjecture \ref{con:plantedclique} holds for some constants $\alpha, \delta \in (0, \frac{1}{2})$, and assume that Conjecture \ref{con:sexpander} holds for some constant $\gamma \in (\frac{1}{2}, 1)$. Then for all large enough $\lambda$, the following is true. Set $\bar{n} = 2^{(\log\lambda)^{1/(1+\alpha)}}$, $\bar{k} = \bar{n}^\delta$, $\ell=(\log\bar{n})^\alpha$, $q=\bar{n}^{-(\delta\ell)/2}$, $\zeta = \lceil\frac{1}{1-\sqrt{\gamma}}\rceil + 1$, $\beta = \zeta \cdot (\zeta(1-\alpha)+2)$, and $\rho=2+\beta\zeta$. For all \text{poly}$(\lambda)$ size algorithms $\cA$,
        \begin{align*}
            & \Big \lvert \Pr \left[ (\pk, \sk) \from \Gen(1^\lambda); \cA(\pk, \Enc(\pk, 0)) = 1,  \right ]\\
            & \qquad - \Pr \left[ (\pk, \sk) \from \Gen(1^\lambda);  \cA(\pk, \Enc(\pk, 1)) = 1 \right ] \Big \rvert \leq 0.99,
        \end{align*}
        where the probability is taken over the coins of $\Encode$, $\Gen$, $\Enc$, and $\cA$.
\end{lemma}

\begin{proof}
    We will use standard hybrid arguments.
    \begin{mdframed}
        Hybrid $\cH_0(1^\lambda)$:
    \begin{enumerate}
        \item $G \getsr \cG(\bar{n},\frac{1}{2},\bar{k})$.
        \item $H \getsr \RP^\ell_p(G)$.
        \item $\mat{S} \gets \Encode(H,\zeta,\beta)$. Let $\mat A$ have $m$ rows and $n$ columns.
        \item Let $\mathbf{A} \in \mathbb{F}_p^{m \times n}$ be the matrix where we replace each nonzero entry in $\mathbf{S}$ with a random element in $\mathbb{F}_p^*$.
        \item $\pk \coloneqq \mat{A}$.
        \item $\ct \from \Enc(\pk, 0)$ where
        \[\ct \xleftarrow{} (\mat y, \mat{A}\mat{s}+\mat{e})\]
        Here $\mat{s} \getsr \bF_p^{n}$, and then we sample $\mat{y} \getsr \bF_2^{n \lceil\log p\rceil}$ conditioned on $\mathbf{\tilde{s}} \cdot \mat{y} = 0$. We sample $\mathbf{e}$ as:
        \[\mathbf{e}_i=\begin{cases} \text{Random element of $\mathbb{F}_{p}^*$, with probability $\eta= (\log n)^{-\rho}$. } \\ 0, \text{ otherwise.} \end{cases}\]
        \item Output $(\pk, \ct)$.
    \end{enumerate}
    \end{mdframed}

    \begin{mdframed}
        Hybrid $\cH_{0, \$}(1^\lambda)$:
    \begin{enumerate}
        \item \textcolor{orange}{\underline{$G \getsr \cG(\bar{n},\frac{1}{2})$.}}
        \item $H \getsr \RP^\ell_p(G)$.
        \item $\mat{S} \gets \Encode(H,\zeta,\beta)$. Let $\mat A$ have $m$ rows and $n$ columns.
        \item Let $\mathbf{A} \in \mathbb{F}_p^{m \times n}$ be the matrix where we replace each nonzero entry in $\mathbf{S}$ with a random element in $\mathbb{F}_p^*$.
        \item $\pk \coloneqq \mat{A}$.
        \item $\ct \from \Enc(\pk, 0)$ where
        \[\ct \xleftarrow{} (\mat y, \mat{A}\mat{s}+\mat{e})\]
        Here $\mat{s} \getsr \bF_p^{n}$, and then we sample $\mat{y} \getsr \bF_2^{n \lceil\log p\rceil}$ conditioned on $\mathbf{\tilde{s}} \cdot \mat{y} = 0$. We sample $\mathbf{e}$ as:
        \[\mathbf{e}_i=\begin{cases} \text{Random element of $\mathbb{F}_{p}^*$, with probability $\eta= (\log n)^{-\rho}$. } \\ 0, \text{ otherwise.} \end{cases}\]
        \item Output $(\pk, \ct)$.
    \end{enumerate}
    \end{mdframed}

    \begin{mdframed}
        Hybrid $\cH_{1, \$}(1^\lambda)$:
    \begin{enumerate}
        \item $G \getsr \cG(\bar{n},\frac{1}{2})$.
        \item $H \getsr \RP^\ell_p(G)$.
        \item $\mat{S} \gets \Encode(H,\zeta,\beta)$. Let $\mat A$ have $m$ rows and $n$ columns.
        \item Let $\mathbf{A} \in \mathbb{F}_p^{m \times n}$ be the matrix where we replace each nonzero entry in $\mathbf{S}$ with a random element in $\mathbb{F}_p^*$.
        \item $\pk \coloneqq \mat{A}$.
        \item $\ct \from \Enc(\pk, 0)$ where
        \[\ct \xleftarrow{} (\mat y, \mat{A}\mat{s}+\mat{e})\]
        \textcolor{orange}{\underline{Here $\mat{s} \getsr \bF_p^{n}$, and then we sample $\mat{y} \getsr \bF_2^{n \lceil\log p\rceil}$ conditioned on $\mathbf{\tilde{s}} \cdot \mat{y} = 1$.}} We sample $\mathbf{e}$ as:
        \[\mathbf{e}_i=\begin{cases} \text{Random element of $\mathbb{F}_{p}^*$, with probability $\eta= (\log n)^{-\rho}$. } \\ 0, \text{ otherwise.} \end{cases}\]
        \item Output $(\pk, \ct)$.
    \end{enumerate}
    \end{mdframed}

    \begin{mdframed}
        Hybrid $\cH_{1}(1^\lambda)$:
    \begin{enumerate}
        \item \textcolor{orange}{\underline{$G \getsr \cG(\bar{n},\frac{1}{2},\bar{k})$.}}
        \item $H \getsr \RP^\ell_p(G)$.
        \item $\mat{S} \gets \Encode(H,\zeta,\beta)$. Let $\mat A$ have $m$ rows and $n$ columns.
        \item Let $\mathbf{A} \in \mathbb{F}_p^{m \times n}$ be the matrix where we replace each nonzero entry in $\mathbf{S}$ with a random element in $\mathbb{F}_p^*$.
        \item $\pk \coloneqq \mat{A}$.
        \item $\ct \from \Enc(\pk, 0)$ where
        \[\ct \xleftarrow{} (\mat y, \mat{A}\mat{s}+\mat{e})\]
        Here $\mat{s} \getsr \bF_p^{n}$, and then we sample $\mat{y} \getsr \bF_2^{n \lceil\log p\rceil}$ conditioned on $\mathbf{\tilde{s}} \cdot \mat{y} = 1$. We sample $\mathbf{e}$ as:
        \[\mathbf{e}_i=\begin{cases} \text{Random element of $\mathbb{F}_{p}^*$, with probability $\eta= (\log n)^{-\rho}$. } \\ 0, \text{ otherwise.} \end{cases}\]
        \item Output $(\pk, \ct)$.
    \end{enumerate}
    \end{mdframed}

    As before, before analyzing the hybrids, note that $\lambda = \bar{n}^{\Theta(\ell)}$ and $\lambda = n^{\Theta(1)}$ with very high probability. \\\\
    We first argue that $\cH_0 \approx_{\frac{1}{3}} \cH_{0, \$}$ and $\cH_{1, \$} \approx_{\frac{1}{3}} \cH_{1}$, which follows from the planted clique conjecture.
    
    \begin{claim}[$\cH_0 \approx_{\frac{1}{3}} \cH_{0, \$}$] \label{claim:clique1}
        Assume that Conjecture \ref{con:plantedclique} holds for some constants $\alpha, \delta \in (0, \frac{1}{2})$. For all sufficiently large $\lambda$, and for all constant $\gamma \in (\frac{1}{2}, 1)$, the following is true. Set the parameters as in the lemma statement. For all poly$(\lambda)$-size algorithms $\cA$,
        \[\Abs{\Pr \left[ \cA (1^\lambda,\cH_0(1^\lambda)) = 1 \right] - \Pr \left[ \cA (1^\lambda, \cH_{0, \$}(1^\lambda)) = 1 \right] } \leq \frac{1}{3}.\]
    \end{claim}
    \begin{proof}
        Identical to the proof of Claim~\ref{lemma:h0h0r}.
    \end{proof}
    \begin{claim}[$\cH_{1, \$} \approx_{\frac{1}{3}} \cH_{1}$] \label{claim:clique2}
        Assume that Conjecture \ref{con:plantedclique} holds for some constants $\alpha, \delta \in (0, \frac{1}{2})$. For all sufficiently large $\lambda$, and for all constant $\gamma \in (\frac{1}{2}, 1)$, the following is true. Set the parameters as in the lemma statement. For all poly$(\lambda)$-size algorithms $\cA$,
        \[\Abs{\Pr \left[ \cA (1^\lambda,\cH_{1,\$}(1^\lambda)) = 1 \right] - \Pr \left[ \cA (1^\lambda, \cH_{1}(1^\lambda)) = 1 \right] } \leq \frac{1}{3}.\]
    \end{claim}
    \begin{proof}
        Identical to the proof of Claim~\ref{lemma:h1rh1}.
    \end{proof}
    We now argue that $\cH_{0, \$} \approx_{0.30} \cH_{1, \$}$ via an appeal to the Goldreich-Levin Theorem~\cite{goldreich89}. We first state a version of this theorem below:
    \begin{claim}[Adapted from Theorem 3 in \cite{bellare99}]
        \label{claim:GLhelper}
        Suppose we have an unknown vector $\mat s \in \mathbb{F}_2^n$, but we know the value of $f(\mat s)$ for some function $f$. Suppose further that we have a poly$(n)$-size algorithm $\cQ$ satisfying
        \[\Pr_{\mat y \getsr \mathbb{F}_2^n}[\cQ(\mat y, f(\mat s)) = \mat y \cdot \mat s] \geq 0.51,\]
        and a poly$(n)$-size algorithm $\cY$ satisfying
        \[\cY(\mat s', f(\mat s)) = 1 \textnormal{ if and only if } \mat s' = \mat s.\]
        Then there exists a poly$(n)$-size algorithm $\cZ$ such that
        \[\Pr[\cZ(f(\mat s)) = \mat s] = 1 - o(1).\]
    \end{claim}
    
    \begin{claim}[$\cH_{0, \$} \approx_{0.30} \cH_{1, \$}$] \label{claim:GL}
        Assume that Conjecture \ref{con:sexpander} holds for some constant $\gamma \in (\frac{1}{2}, 1)$. For all sufficiently large $\lambda$, and for all constants $\alpha, \delta \in (0, \frac{1}{2})$, the following is true. Set the parameters as in the lemma statement. For all poly$(\lambda)$-size algorithms $\cA$,
        \[\Abs{\Pr \left[ \cA (1^\lambda,\cH_{0,\$}(1^\lambda)) = 1 \right] - \Pr \left[ \cA (1^\lambda, \cH_{1, \$}(1^\lambda)) = 1 \right] } \leq 0.30.\]
    \end{claim}

    \begin{proof}
    Towards a proof by contradiction, suppose there exists a poly$(\lambda)$-size algorithm $\cA$ to differentiate between $\cH_{0, \$}$ and $\cH_{1, \$}$ with advantage more than $0.30$. By randomly sampling $\lambda$ many choices of $(\mat A, \mat s, \mat e)$ as generated in hybrids $\cH_{0, \$}$ and $\cH_{1, \$}$, and then executing $\cA$ on these, we can approximate
    \[p_0 \approx \Pr \left[ \cA (1^\lambda,\cH_{0,\$}(1^\lambda)) = 1 \right] \textnormal{ and}\]
    \[p_1 \approx \Pr \left[ \cA (1^\lambda, \cH_{1, \$}(1^\lambda)) = 1 \right]\]
    to within an error of $\pm o(1)$, with probability $1 - o(1)$.\\\\
    Suppose the above procedure indicates that $p_1 > p_0$. Then we define an algorithm $\cB$ which (i) takes as input a tuple $(\mat y, \mat A, \mat A \mat s + \mat e)$, (ii) gives the tuple as input to algorithm $\cA$, and (iii) outputs 1 if $\cA$ outputs 1, and outputs 0 otherwise. If instead the procedure indicates that $p_1 < p_0$, we define $\cB$ identically, except it outputs 1 if $\cA$ does \emph{not} output 1, and it outputs 0 otherwise. Assuming the estimates for $p_0$ and $p_1$ are accurate to within $\pm o(1)$, we have that
    \begin{align}
        & \Pr_{(\mat y, \mat A, \mat s, \mat e) \textnormal{ is sampled as in } \cH_{1, \$}}[\cB(\mat y, \mat A, \mat A \mat s + \mat e) = 1] - \Pr_{(\mat y, \mat A, \mat s, \mat e) \textnormal{ is sampled as in } \cH_{0, \$}}[\cB(\mat y, \mat A, \mat A \mat s + \mat e) = 1] > 0.30.
    \end{align}
    This ensures that, with high probability, $\cB$ is more likely to return 1 when the dot product $\mat y \cdot \mat{\tilde{s}}$ does indeed equal 1 (the original algorithm $\cA$ may have the reverse property). \\\\
    We now perform a probabilistic analysis of $\cB$. Define
    \begin{align*}
        f(\mat A, \mat s, \mat e) = & \Pr_{\mat y \getsr \mathbb{F}_p^n \textnormal{ conditioned on } \mat{\tilde{s}} \cdot \mat y = 1}[\cB(\mat y, \mat A, \mat A \mat s + \mat e) = 1] - \Pr_{\mat y \getsr \mathbb{F}_p^n \textnormal{ conditioned on } \mat{\tilde{s}} \cdot \mat y = 0}[\cB(\mat y, \mat A, \mat A \mat s + \mat e) = 1].
    \end{align*}
    Note that $f(\mat A, \mat s, \mat e) \in [-1, 1]$ for all $(\mat A, \mat s, \mat e)$. The sampling procedure for $(\mat A, \mat s, \mat e)$ is the same in both hybrids, so we can rewrite (1) as
    \begin{align}
        \sum_{(\mat A, \mat s, \mat e)}\Pr[(\mat A, \mat s, \mat e) \textnormal{ is sampled in } \cH_{1, \$}] \cdot f(\mat A, \mat s, \mat e) > 0.30.
    \end{align}
    Let $\mathcal{T}$ be the set of all possible $(\mat A, \mat s, \mat e)$, and let $\mathcal{T}_{\geq \varepsilon} \subseteq \mathcal{T}$ be the set of all $(\mat A, \mat s, \mat e)$ such that $f(\mat A, \mat s, \mat e) \geq \varepsilon$. For all $(\mat A, \mat s, \mat e) \in \mathcal{T}\backslash\mathcal{T}_{\geq \varepsilon}$, we have the trivial upper bound that $f(\mat A, \mat s, \mat e) < \varepsilon$. For all $(\mat A, \mat s, \mat e) \in \mathcal{T}_{\geq \varepsilon}$, we have the trivial upper bound that $f(\mat A, \mat s, \mat e) \leq 1$. This combined with (2) implies that
    \begin{align*}
        & \sum_{(\mat A, \mat s, \mat e) \in \mathcal{T}_{\geq \varepsilon}}\Pr[(\mat A, \mat s, \mat e) \textnormal{ is sampled in } \cH_{1, \$}] \\
        & \qquad + \varepsilon \cdot \sum_{(\mat A, \mat s, \mat e) \in \mathcal{T}\backslash\mathcal{T}_{\geq \varepsilon}}\Pr[(\mat A, \mat s, \mat e) \textnormal{ is sampled in } \cH_{1, \$}] > 0.30,
    \end{align*}
    and hence,
    \begin{align*}
        & \sum_{(\mat A, \mat s, \mat e) \in \mathcal{T}_{\geq \varepsilon}}\Pr[(\mat A, \mat s, \mat e) \textnormal{ is sampled in } \cH_{1, \$}] > 0.30 - \varepsilon.
    \end{align*}
    Expressed equivalently, we have
    \begin{align*}
        & \Pr_{(\mat A, \mat s, \mat e) \textnormal{ is sampled as in } \cH_{1, \$}}[(\mat A, \mat s, \mat e) \in \mathcal{T}_{\geq \varepsilon}] > 0.30 - \varepsilon.
    \end{align*}
    This is useful because, for all $(\mat A, \mat s, \mat e) \in \mathcal{T}_{\geq \varepsilon}$, by the definition of $\mathcal{T}_{\geq \varepsilon}$, we have the \emph{lower} bound that
    \begin{align}
        \Pr_{\mat y \getsr \mathbb{F}_p^n \textnormal{ conditioned on } \mat{\tilde{s}} \cdot \mat y = 1}[\cB(\mat y, \mat A, \mat A \mat s + \mat e) = 1] - \Pr_{\mat y \getsr \mathbb{F}_p^n \textnormal{ conditioned on } \mat{\tilde{s}} \cdot \mat y = 0}[\cB(\mat y, \mat A, \mat A \mat s + \mat e) = 1] \geq \varepsilon.
    \end{align}
    This analysis actually shows that we can interpret $\cB$ as a predictor for the dot product $\mat y \cdot \mat s$. (This will serve as algorithm $\cQ$ when we apply the Goldreich-Levin Theorem.) To see this, first let
    \begin{align}
        p_{\textnormal{avg}} = \frac{1}{2} \cdot \left(\Pr_{\mat y \getsr \mathbb{F}_p^n \textnormal{ conditioned on } \mat{\tilde{s}} \cdot \mat y = 1}[\cB(\mat y, \mat A, \mat A \mat s + \mat e) = 1] + \Pr_{\mat y \getsr \mathbb{F}_p^n \textnormal{ conditioned on } \mat{\tilde{s}} \cdot \mat y = 0}[\cB(\mat y, \mat A, \mat A \mat s + \mat e) = 1]\right).
    \end{align}
    Now we analyze the behavior of $\cB$ on $(\mat y, \mat A, \mat A \mat s + \mat e)$, where $\mat y$ is a random vector $\mat y \getsr \mathbb{F}_2^n$ and we assume $(\mat A, \mat s, \mat e) \in \mathcal{T}_{\geq \varepsilon}$. With probability $\frac{1}{2}$ over the choice of $\mat y$, we have that $\mat y \cdot \mat s = 1$, and in this case
    \[\Pr[\cB(\mat y, \mat A, \mat A \mat s + \mat e) = 1 = \mat y \cdot \mat{\tilde{s}}] \geq p_{\textnormal{avg}} + \frac{\varepsilon}{2}.\]
    With the remaining probability $\frac{1}{2}$ over the choice of $\mat y$, we have $\mat y \cdot \mat{\tilde{s}} = 0$, and in this case
    \[\Pr[\cB(\mat y, \mat A, \mat A \mat s + \mat e) = 1] \leq p_{\textnormal{avg}} - \frac{\varepsilon}{2}.\]
    which implies
    \[\Pr[\cB(\mat y, \mat A, \mat A \mat s + \mat e) = 0 = \mat y \cdot \mat s] \geq 1 - p_{\textnormal{avg}} + \frac{\varepsilon}{2}.\]
    Accumulating these probabilities, we have the following when $\mat y \getsr \mathbb{F}_2^n$ and $(\mat A, \mat s, \mat e) \in \mathcal{T}_{\geq \varepsilon}$:
    \[\Pr[\cB(\mat y, \mat A, \mat A \mat s + \mat e) = \mat y \cdot \mat s] \geq \frac{1}{2}\cdot (p_{\textnormal{avg}} + \frac{\varepsilon}{2}) + \frac{1}{2} \cdot (1 - p_{\textnormal{avg}} + \frac{\varepsilon}{2}) = \frac{1}{2} + \frac{\varepsilon}{2}.\]
    If we take $\varepsilon = 0.04$, then the predictor has accuracy $0.52 > 0.51$.
    \\\\
    We can also define a poly$(\lambda)$-size algorithm $\mathcal{D}(\mat s', \mat A, \mat A \mat s + \mat e)$ such that, with $1 - o(1)$ probability over the choice of $(\mat A, \mat s, \mat e)$ as produced in $\cH_{0, \$}$ and $\cH_{1, \$}$, we have that $\mathcal{D}(\mat s', \mat A, \mat A \mat s + \mat e) = 1$ if and only if $\mat s' = \mat s$. (Note that we can freely convert between the bit representation $\mathbf{\tilde{s'}}$ and the original vector $\mathbf{s'}$.) This will serve as algorithm $\cY$ when we apply the Goldreich-Levin Theorem. The algorithm will simply compute $\mat b' = \mat A \mat s'$, determine the Hamming distance between $\mat b'$ and $\mat b = \mat A \mat s + \mat e$, and then return either 0 or 1 based on the magnitude of the Hamming distance. With $1 - o(1)$ probability over the choice of $(\mat A, \mat s, \mat e)$ as produced in $\cH_{0, \$}$ and $\cH_{1, \$}$, if we set $\mat s' = \mat s$, the Hamming distance will be significantly smaller than for all other choices of $\mat s'$. (This property is folklore. Roughly speaking, it follows because, when $\zeta$ is a constant, every column of $\mat A$ will have polynomial Hamming weight with very high probability.) We can determine an appropriate Hamming distance cutoff in polynomial time. \\\\
    Now we apply Claim~\ref{claim:GLhelper} to get a poly$(\lambda)$-size algorithm $\mathcal{X}$ that recovers $\mat s$ from  $f(\mat s)$, where $f(\mat s) = (\mat A, \mat A \mat s + \mat e)$. The algorithm follows by simply applying the claim with $\cQ := \cB$, $\cY := \cD$, and $\cX := \cZ$.\\\\
    Assuming that $(\mat A, \mat s, \mat e) \in \mathcal{T}_{\geq 0.04}$, that we can construct the algorithm $\cD$, and that our estimates for $p_0, p_1$ are accurate to within $\pm o(1)$, the algorithm $\cX$ will recover the vector $\mat s$ with $1 - o(1)$ probability. Applying the probability bounds we derived above, we see that $\cX$ recovers $\mat s$ from $(\mat A, \mat A \mat s + \mat e)$ with probability $0.26 - o(1)$ when $(\mat A, \mat s, \mat e)$ is sampled as in $\cH_{0, \$}$ and $\cH_{1, \$}$. \\\\
    As in the proof of Claim~\ref{lemma:h0rh1r}, the matrix $\mat{A}$ will satisfy the preconditions of Conjecture~\ref{con:sexpander} with $1 - o(1)$ probability. Furthermore, the sampling procedure for $\mat s$ and $\mat e$ is the same in the statement of Conjecture~\ref{con:sexpander} as in the hybrids $\cH_{0, \$}$ and $\cH_{1, \$}$. Combining this with the above analysis of algorithm $\cX$, we see that $\cX$ is an algorithm for the search problem in Conjecture~\ref{con:sexpander} that succeeds with probability at least $0.26 - o(1) > \frac{1}{4}$, which is a contradiction.
    \end{proof}
    
    Combining Claims~\ref{claim:clique1},~\ref{claim:clique2},~and~\ref{claim:GL}, and accounting for the fact that $\lambda = \bar{n}^{\Theta(\ell)}$ and $\lambda = n^{\Theta(1)}$ with very high probability, we get  \begin{align*}
            & \Big \lvert \Pr \left[ (\pk, \sk) \from \Gen(1^\lambda); \cA(\pk, \Enc(\pk, 0)) = 1,  \right ]\\
            & \qquad - \Pr \left[ (\pk, \sk) \from \Gen(1^\lambda);  \cA(\pk, \Enc(\pk, 1)) = 1 \right ] \Big \rvert \leq \frac{29}{30} + o(1) < 0.99.
        \end{align*} 
\end{proof}

\begin{theorem}\label{thm:searchscheme}
    Assuming Conjecture \ref{con:plantedclique} and Conjecture \ref{con:sexpander} both hold, there exists a semantically secure public key encryption scheme.
\end{theorem}

\begin{proof}
    Identical to the proof of Theorem \ref{thm:scheme}.
\end{proof}